\documentstyle[epsfig]{mn}

\newcommand{\msun}{\mbox{$M_{\odot}$}}
\newcommand{\etal}{\mbox{et al.\ }}
\newcommand{\ie}{\mbox{i.e.,\ }}
\newcommand{\eg}{\mbox{e.g.,\ }}

\newcommand{\Lone}{$L^{(1)}\:$}
\newcommand{\fullcat}{\mbox{77\ }}
\newcommand{\cutcat}{\mbox{52\ }}
\newcommand{\stdcat}{\mbox{34\ }}
\newcommand{\exccat}{\mbox{25\ }}
 
\begin{document}
 

\title[Linear-theory predictions of galaxy formation]
 {Testing linear-theory predictions of galaxy formation}
\author[B. Sugerman, F.J. Summers, M. Kamionkowski]
 {Ben Sugerman,$^{1,2{\textstyle \star}}$
  F. J. Summers,$^{2,3{\textstyle \star}}$
  and Marc Kamionkowski$^{2,4}$\thanks{
   E-mail: {ben@astro.columbia.edu (BS)}; {summers@amnh.org (FJS)};
   {kamion@phys.columbia.edu (MK)}
        }\\
  $^1$Department of Astronomy, Columbia University, 538 West 120th
 Street, New York, NY 10027, USA\\
  $^2$Columbia Astrophysics Lab, Columbia University, 538 West 120th
 Street, New York, NY 10027, USA\\
  $^3$Department of Astrophysics, American Museum of Natural History,
 Central Park West at 79$^{\rm th}$ St., New York, NY 10024, USA\\
  $^4$Department of Physics, Columbia University, 538 West 120th
 Street, New York, NY 10027, USA}
\maketitle 

\begin{abstract}

The angular momentum of galaxies is routinely ascribed to a process of
tidal torques acting during the early stages of gravitational
collapse, and is predicted from the initial mass distribution using
second-order perturbation theory and the Zel'dovich approximation.
We have tested this theory for a flat hierarchical cosmogony using a
large $N$-body simulation with sufficient dynamic
range to include tidal fields, allow resolution of
individual galaxies, and thereby expand on previous studies. 
The predictions of linear collapse, linear tidal torque, and
biased-peaks galaxy formation are applied to the initial conditions
and compared to evolved bound objects. We find relatively good
correlation between the predictions of linear theory and actual galaxy
evolution.  Collapse is well described by an ellipsoidal model within
a shear field, which results primarily in triaxial objects which do
not map directly to the initial density field.  While structure
formation from early times is a complex history of hierarchical
merging, salient features are well described by the simple
spherical-collapse model.  Most notably, we test several methods for
determining the turnaround epoch, and find that turnaround is
succesfully described by the spherical collapse model.  
The angular momentum of collapsing
structures grows linearly until turnaround, as predicted, and
continues quasi-linearly until shell crossing.  The predicted angular
momentum for well-resolved galaxies at turnaround overestimates the
true turnaround and final values by a factor of $\sim 3$ with a
scatter of $\sim 70 $ percent, and only marginally yields the correct
direction of the angular momentum vector.  We recover the prediction
that final angular momentum scales as mass to the 5/3 power.  We find
that mass and angular 
momentum also vary proportionally with peak height.  In view of the
fact that the observed galaxy collapse is a stochastic hierarchical
and non-linear process, it is encouraging that the linear theory can
serve as an effective predictive and analytic tool.  

\end{abstract}

\begin{keywords} cosmology -- dark matter -- galaxies:formation --
numerical methods -- large scale structure of Universe
 \end{keywords}
 
\section{Introduction}

Large-scale-structure formation (\ie galaxies and clusters)
most likely resulted from gravitational
amplification of small density inhomogeneities in an otherwise smooth
primordial density field following an initial hot big-bang. This
scenario is supported by the {\it COBE} findings of the cosmic
microwave background radiation anisotropies (Smoot \etal 1992).  In
the standard-CDM (cold dark matter) model (Peebles 1982; Bond \&
Szalay 1983; Bond \& Efstathiou 1984), these fluctuations grow in the
non-baryonic dark-matter component before recombination,
thereby providing the density perturbations which evolve into
structures and voids (for reviews, see Ostriker 1993; Peebles 1993).
The evolution of these inhomogeneities are relatively easily studied
using linear perturbation theory as long as the density contrast is
small compared to unity, \ie within the linear regime.  However, once
the density contrast exceeds unity, linear theory breaks down and
either analytic approximations or $N$-body simulations must be invoked
to study the evolution of perturbations.   

Application of perturbation approximations in the linear
regime may offer insight into a protogalaxy's acquisition of angular
momentum during separation and collapse from the Hubble flow.
Hoyle (1949) first suggested that a protogalaxy's spin arises from the
tidal fields of its neighbors.  An alternative theory, in which galactic
spin is the relic of primeval vorticity, is not widely regarded as
favorable (see the review by Efstathiou \& Silk 1983).  Peebles (1969)
first examined Hoyle's theory using a second-order perturbative
expansion to  find the growth rate of angular momentum induced by
gravitational tidal torques from the surrounding expanding
matter field.  In particular, he found that the angular momentum
within a spherical comoving volume grows as $t^{5/3}$ in an
Einstein-de Sitter universe. Second-order expansion is necessary,
since to first order, an isotropic spherical volume can not gain
angular momentum from external torques.  Doroshkevich (1970) found
that for a generic non-spherical volume enclosing a protogalaxy, the 
angular-momentum growth rate at early times (in a flat
cosmology) is linearly proportional to time, and that
Peebles' growth rate resulted from incorrect symmetries.
White (1984) determined that Peebles' findings
resulted from surface effects which convect angular momentum across
the boundary of the enclosing Eulerian volume, and correctly
reinterpreted the analysis to yield the predicted linear growth
rate.  White noted that his approximations strictly apply only to early
times, during which the density contrast (or the density field
convolved with a window function of protogalactic scale) is much less
than unity, and that the final spin is dependent entirely on how the
tidal torquing is terminated.  Following the collapse history dictated
by the spherical-collapse model (Partridge \& Peebles 1967; Gunn \&
Gott 1972; Peebles 1980 \S19), one typically assumes that a galaxy
becomes insensitive to the external tidal field once it separates from
the Hubble flow and begins collapse (Peebles 1969).    

Peebles' and White's analyses involve volumes
centred on random points, hence the matter contained within those
volumes is not guaranteed to collapse into a bound protogalaxy.
However, we expect that protogalaxies will form in regions of density
peaks, currently known as the biased galaxy-formation model (\eg
Kaiser 1984; Peacock \& Heavens 1985; Bardeen \etal 1986).  A revised
model has been proposed, in which  
second-order perturbation expansions describe tidal torques from
high-density peaks acting on the primordial mass distribution of a
protogalaxy, and has been analyzed by, for example, Hoffman (1986,
1988), and Catelan \& Theuns (1996a, hereafter CT96).  The acquisition
of angular momentum from tidal fields and the role of tidal shear has
also been studied by, e.g.\ Binney \& Silk 1979; Heavens \& Peacock
(1988, hereafter HP88); Ryden (1988); Quinn \& Binney (1992); Dubinski
(1992); Zaroubi 
\& Hoffman (1993); Bond \& Myers (1993); Bertschinger \& Jain (1994);
van de Weygaert \& Babul (1994); and Eisenstein \& Loeb (1995, hereafter
EL95).   

Early quantitative comparison of angular-momentum evolution with
linear theory was first addressed by Peebles (1971) and
Efstathiou \& Jones (1979).  White (1984), using two $32,\!768$ particle
$N$-body simulations, found that the {\em mean} angular momentum over
all groups grows linearly at very early times ($a \la 2$),
however from the individual examples he shows, one is led to conclude
that the dispersion about this value is quite high.  Barnes \&
Efstathiou (1987, hereafter BE87) also draw the
same conclusion from 7 simulations using $8,\!000-32,\!768$ particles and
a variety of initial conditions.  Again, the mean growth of $L$ at
early times appears linear, while the individual galaxies
exhibit a significant dispersion.  BE87 further concluded that the
linear prediction for the final angular momentum provides only an
order-of-magnitude estimate.  

In this paper, we address the predictions of linear theory and test
their validity using a P3MSPH gravitational and hydrodynamic
simulation.  We expand on previous numerical studies by using galaxies
evolved with higher numerical and spatial resolution.
Following the collapse history and evolution of \fullcat
galaxies from high redshift until the present epoch, we compare the
angular momentum to linear theory, to study the accuracy and limits of
applicability of this formalism.  We address whether the linear-theory
approximation is applicable in the weakly non-linear and strictly
non-linear regimes.  We examine the role and significance
of local effects such as mergers and close tidal encounters on the
evolutionary history. We further examine various models of
protogalactic collapse and comment on the applicability and limitations
of each.   In a
hierarchical-clustering cosmogony, structure formation is not expected
to follow a simple spherical-collapse model.  Therefore, it is most
advantageous if the linear theory successfully predicts the final
angular momentum of a collapsing protogalaxy, for use as a
computational tool, numerical shortcut, and in reconstructing the
primordial density field.

The layout of the paper is as follows.  
In \S 2, we briefly review White's formalism for linear tidal
torques, as well as the predictions of the spherical-collapse model.
We describe the numerical simulation, group-finding, and our galaxy
catalog in \S 3, and our methodology for measuring 
pertinent parameters for each galaxy in \S 4.  Results are presented
in \S 5.  We first examine the collapse history of galaxies and
compare the qualitative and quantitative predictions of linear theory
to the evolution of our set of galaxies. We then discuss scalings
derived from linear theory, compare predictions of the biased
galaxy-formation scenario to our data, and briefly discuss
correlations.  The paper concludes with a discussion in \S6.

\section{Dynamical formalism \label{sec2}}

For completeness, we present the derivation of the linear theory
developed by White (1984) using the Zel'dovich (1970) approximation.
Consider the matter within an expanding Friedmann universe to be
adequately described as a Newtonian pressureless cold fluid.
The linear-theory mapping between the Eulerian and Lagrangian
comoving coordinates ${\bmath x}$ and ${\bmath q}$
(respectively) is given by \begin{equation}
 {\bmath x}({\bmath q},t)={\bmath q}-
 D(t) \nabla \psi({\bmath q}), \label{eqn2-0}
\end{equation}
where ${\bmath q}$ is defined as
the ${\bmath x}$ position of a particle as $t \rightarrow 0$;
and $\nabla \psi$ is proportional to the peculiar gravitational
potential.  $D(t)$ describes the growing density mode.
Equation (\ref{eqn2-0}) is a statement of the Zel'dovich (1970)
approximation,  and is strictly valid only if $\langle\delta\rangle
\ll 1$,  where the fractional density contrast
$\delta({\bmath r},t)= \rho({\bmath r},t)/\rho_{b}(t) - 1$.
This condition is satisfied if the initial
density fluctuations have coherence lengths of protogalactic
size. However since this may only be the case at the earliest times, 
we smooth the density field by convolving it with a window function of
protogalactic scale.  This implicitly assumes that small-scale
nonlinearities have negligible effect on large-scale
fluctuations at early times (White 1984).  

The angular momentum of the matter contained within an Eulerian volume
$V$, neglecting centre of mass motion, is given by \begin{equation}
  {\bmath L}(t)=\int_{a^{3}V}d{\bmath r}\rho
  {\bmath r \times \dot{r}} = \rho_{b}a^{5}\int_{V}
  d{\bmath x}(1+\delta){\bmath x \times u},
  \label{eqn2-1}
\end{equation}
where ${\bmath u}=\dot{\bmath x}$; a dot refers to the normal time
derivative; ${\bmath r} = a(t)
{\bmath x}$; and $a$ is the cosmological expansion 
factor.  If the transformation tensor 
{\sffamily\bfseries T}
between Eulerian and Lagrangian coordinates is
defined such that ${\bmath x(q},t)=\mbox{{\sffamily\bfseries T}}{\bmath q}$, then 
the Jacobian $J$ of {\sffamily\bfseries T} is the change in volume
between Eulerian  and Lagrangian coordinates.
Since Eulerian and Lagrangian coordinates are
identical at $t=0$, $J(t=0)\equiv 1$, it follows directly that 
${\rho}/{\rho_{0}} = {J(0)}/{J(t)}$.  If 
$\rho_{b}=\rho_{0}$, \ie the background mean density is equal to the
primordial density, then we immediately retrieve the continuity
equation,
\begin{equation}
  1+\delta[{\bmath x}({\bmath q},t),t] =J^{-1}. \label{eqn2-2}
\end{equation}

From equation (\ref{eqn2-0}), we find the peculiar velocity,
\begin{equation}
 {\bmath v}({\bmath q},t)= a {\bmath u}({\bmath q},t)
  =a\dot{D}\nabla \psi. \label{eqn2-3}
\end{equation}
We now transform the integral (\ref{eqn2-1}) to the Lagrangian volume
$\Gamma$,
\begin{equation}
  {\bmath L}(t)
  =\eta_{0}a^{2}\dot{D}\int_{\Gamma}d{\bmath q} \, {\bmath q}
  \times \nabla\psi({\bmath q}), \label{eqn2-4}
\end{equation} 
where $\eta_{0}\equiv\rho_{b}a^{3}=\rho_{0}a_{0}^{3}$. It is obvious
from equation (\ref{eqn2-4}) that the angular momentum ${\bmath L}$
depends on the shape of the volume $\Gamma$ that encloses the
constituents of the protogalaxy.   Following White (1984), we
approximate the potential $\psi$ by the first three terms of the
Taylor series about the origin ${\bmath q=0}$.  The linear
approximation for the angular momentum may then be expressed as 
\begin{equation}
  L^{(1)}_{i}(t)=a(t)^{2}\dot{D}(t)\epsilon_{ijk}
  {\cal D}^{(1)}_{jl}{\cal I}_{lk} \label{eqn2-5}
\end{equation}
where
\begin{equation}
 {\cal D}^{(1)}_{ij}=\partial_{i}\partial_{j}\psi
 (\mbox{\boldmath $0$}) \label{eqn2-6}  
\end{equation}
is the initial deformation tensor at the origin ${\bmath q=0}$; 
\begin{equation}
  {\cal I}_{ij}=\eta_{0}\int_{\Gamma} q_{i}q_{j}d{\bmath q}
  \label{eqn2-7}
\end{equation}
is the inertia tensor; $\epsilon_{ijk}$ is the
anti-symmetric Levi-Civita tensor; and summation over repeated indices
is implicit.  Since the principal axes of the deformation and
inertia tensors are, in general, not aligned for a non-spherical
volume, this linear angular momentum should be non-zero.

The time dependence in \Lone lies in the
term $a^{2}(t)\dot{D}(t)$, which for an Einstein-de Sitter universe equals
$\frac{2}{3} t/t_{o}^{2}$ such that \Lone grows linearly in
time. 
One simplifies the computation of the deformation tensor using the
Fourier transform of the potential 
$\tilde{\psi}({\bmath k})$ where ${\bmath k}$ is the comoving
Lagrangian wavevector.  The deformation tensor is now 
\begin{equation}
  {\cal D}^{(1)}_{ij}=-\int{\frac{d{\bmath k}}
  {(2\pi)^{3}}k_{i}k_{j}\tilde{\psi}({\bmath k})
  \tilde{W}(k,R)},  \label{eqn2-8}
\end{equation}
and $\tilde{W}(k,R)$ is the Fourier transform of the smoothing
function $W_{R}({\bmath q})$.
By filtering $\psi$ on scale $R$, we effectively filter out
non-linear mode coupling, thereby restricting our approximations to
linear evolution only.  

One expects that the angular momentum of a galaxy will grow only as
long as the particle distribution is sensitive to large-scale tidal
couplings (Peebles 1969, 1980).  When the density contrast is small,
the gravitational field of a proto-galaxy acts to collapse the object
while the rotation induced by the tidal field of the surrounding
matter opposes this 
contraction.  Under the assumption of spherical evolution, the tidal
influence of surrounding matter is important 
until the density contrast $\delta \sim 1$, at which point one can
consider the proto-galaxy as an isolated collapsing system (Peebles
1980).  

To model its early evolution, we treat the proto-galaxy as a spherical
overdense region in an otherwise flat universe (Partridge \& Peebles
1967; Gunn \& Gott 1972; Peebles 1980 \S19), and evolve the
particles according to the closed Friedmann equations (the
spherical-collapse model). Specifically, inside the overdense region, 
$\rho(r)=\rho > \overline{\rho}$, whereas outside, 
$\rho(r)= \overline{\rho}$ and $\overline{\rho}$ is defined by the
flat Friedmann model.  In this formalism, the galaxy should decouple
from the tidal field at the turnaround epoch $t_{M}$, 
defined as the time of maximum expansion and characterized in this
model by an overdensity $\delta \simeq 4.55$.
Virialization
of the proto-galaxy occurs when the inner mass shells pass through the
centre, at which point time-varying gravitational fields dissipate
energy and relax the system.  This dissipation process is known as
violent relaxation and we refer to these shell crossings through
centre as caustic crossings.  In the approximation that we treat a
non-spherical, non-uniform distribution undergoing hierarchical
collapse as a smoothly evolving 
closed universe, we can also predict (to first order) the time of
caustic crossing from the Friedmann equations as 
$t_{C} \simeq 2t_{M}$.

Had the region not collapsed but expanded linearly, turnaround would
occur when the overdensity (Peebles 1980)
\begin{equation}
 \delta = \frac{3}{20}(6\pi)^{2/3}. \label{eqn2-12}
\end{equation}
Since turnaround occurs in the weakly non-linear
regime, we must apply \Lone beyond its region of strict validity.  It
is therefore of interest to study how far into the nonlinear regime,
if even as far as turnaround, one may successfully apply the linear
prediction.  Furthermore, the Jacobian transformation $J^{-1}$ is
non-vanishing up until caustic crossing (Shandarin \& Zel'dovich
1989).  Therefore we expect that \Lone will be valid at most until
caustic crossing.    

\section{Numerical simulation and galaxy catalog \label{sec3}}
\subsection{Numerical simulation \label{sec3-1}}

To test the linear-theory approximation [equation (\ref{eqn2-5})], we have
performed a high-resolution galaxy-formation simulation using
the P3MSPH code (Evrard 1988; Summers 1993) which incorporates large-
and short-scale gravitational 
forces with smooth-particle hydrodynamics (Hockney \& Eastwood 1981;
Monaghan 1992).
We chose a $16h^{-1}$ Mpc comoving box length with
periodic boundary conditions in a standard-CDM Einstein-de Sitter
universe $(\Omega = 1, \Omega_{b}=0.05, \mbox{$H_{0}=100h$ km s$^{-1}$
Mpc$^{-1}$}$, and $h=0.5)$.  Initial conditions were generated using a
so-called ``glass'' initial distribution (White 1994) and the
Zel'dovich approximation.

To follow galaxy-scale evolution while including larger-scale
tidal fields within a reasonable computational time, our simulation
utilizes  multiple resolution regions.  A simulation with $128^{3}$
gas and $128^{3}$ dark matter (DM) particles was evolved until initial
structures had formed.  
Using the results of this partially evolved simulation, we chose a
cubic subregion 8 $h^{-1}$ Mpc on a side for study.  The subregion was
chosen to contain a mixture of density regions, yet avoid the center
of a rich cluster.
The $258,\!530$ gas and $258,\!530$ dark-matter
particles in this subregion were retained at high resolution.  The
remaining particles outside this subregion were combined and collapsed
into $229,\!933$ ``super''-dark-matter (SDM) particles (\eg Katz
\& White 1993). The gas:DM:SDM
mass ratio is $1:19:160$, with a DM mass of roughly $10^{9}$ \msun$\:$
and a total simulation mass of $2.27\times10^{15}\msun$.

The gravitational softening length for a Plummer-law force was set to
8 kpc (comoving), with the SPH smoothing length limited to a 
minimum of one-third the gravitational softening.
The parameters are appropriate to provide enough resolution for the
formation and evolution of galaxies  (see the note in \S5
regarding the use of gas particles within this region).  The
simulation is evolved from the initial epoch at $z=32$ ($t_{i}=68.8$
Myr) to $z=0$ ($t_{f}=13.04$ Gyr) using 5000 time steps of 2.592 
Myr.  Outputs are generated every 25 steps during the initial
evolution (10 percent) and every 100 steps thereafter.

\subsection{Group-finding algorithm \label{sec3-2}} 

It should be clearly noted that numerical simulations of this sort do
not form ``galaxies,'' {\em per se}.  However, we can identify the
sites of most probable galaxy formation as regions of the highest
overdensity, containing ``halos'' of dark-matter particles inside of
whose potential wells reside dense gaseous components.  In
gravitational simulations of this nature, discrete boundaries between
the virialized galaxies and the outer unbound particles do not exist,
therefore the group finding method must rely on an arbitrary
criterion, such as local overdensity or interparticle spacing.  
The problem of group-finding is therefore non-trivial, and has been
thoroughly examined by, \eg Summers, Davis \& Evrard 1995 and Eisenstein
\& Hut 1998.  Ideally, we would prefer to identify a galaxy as {\em
all} the particles in the simulation which are gravitationally bound
as one structure. However, as this is an $N^{2}$ calculation, it is
computationally far too expensive for practical use.  We chose instead
to define a group as a collection of
$N_{min}$ or more particles all separated by less than a specified
linking length $\eta$, known as the ``friends-of-friends'' (FOF)
approach.  Groups found with this method generally lie within a
minimum overdensity contour given by \begin{equation}
 \delta_{min}=\frac{2\Omega}{(4/3)\pi\eta^{3}}. \label{eqn30}
\end{equation}
Since this is only a minimum contour level, the actual average
overdensity of a galaxy will be many times larger.  
The FOF algorithm considers only interparticle spacing as the criteria
for forming a group, and thereby does
not distinguish between particles which are gravitationally bound to the
system.  As a first-order solution, once a list is generated of all
particles forming a group according to the FOF routine, we
recursively compute the local potential of that ensemble and remove
any unbound particles.  

To identify a galaxy, we generate two independent catalogs of gas
and DM groups found using FOF at the final output ($z=0$).  Gas groups
must contain at least 25 bound particles and DM groups at least 100,
both using a linking length, in units of the mean interparticle
spacing, of $\eta=0.075$, which corresponds to a minimum overdensity
contour $\delta_{min} \sim 1000$.  We then cross-correlate the 
positions of the most-bound particle in each group, throwing out any
group which does not have a corresponding counterpart in the other
species, and create a catalog of galaxies containing both bound
DM and gas components.  With this method, we have identified 98
galaxies within the simulation.  Other galaxies exist within cluster
regions of the simulation, but, if the dark-matter halo is mixed
within the general cluster halo, the galaxy is unsuitable for our
analysis. 

\subsection{Selection criteria and galaxy catalog \label{sec3-3}}

Once the catalog is made, the particles within any galaxy may be
traced back to any epoch, since every particle in the simulation is
labelled with an ID number, which it maintains throughout the entire
run.  Groups whose initial distributions were too close to the
subregion boundary must be excluded from the catalog.  Not only do
these have artificially distorted primordial distributions, but since
they are located at the boundary with the SDM region, these border
groups 
contain an unreasonably large number of SDM particles, which
introduce extraordinary gravitational perturbations to the evolving
galaxy.  Therefore we also disregard any galaxies containing a
significant number ($N_{SDM} \ga 0.1N_{DM}$) of SDM particles.

Galaxies which have undergone large merger events or are still merging
at the final output are tagged as questionable with respect to the
angular-momentum analysis. Angular momentum is not in a stable state 
during merging and gravitational relaxation.  Furthermore, 
in a large merger, particles within the constituent clumps at
earlier times may flow out of the system and hence not be included in
the particle list created at the final output.  This can cause (as we
shall see) a significant amount of angular momentum to flow into or
out of the system at any time and radically distort the measured
evolution.\footnote{This may introduce a bias against
non-linear effects, as noted by the referee.  
In this paper we study the predictive capacity of linear theory in 
environments which should be the most stable, and will present a full
examination of evolutionary histories in paper II.  However, of the 13
merging galaxies in the catalog, we discard only 3.} 

Of the 98 galaxies in our catalog, 12 are discarded for violating
boundary conditions; 3 are discarded, and 10 are tagged as questionable
due to mergers;  and 6 are discarded for containing too many SDM
particles (All ``questionable'' galaxies remain in the catalog).
This leaves a usable catalog of \fullcat galaxies.  

\section{Galaxy parameters \label{sec4}}

For each of the galaxies identified in our catalog, we measure a
variety of characteristics for analysis, outlined in the following
subsection.  

\subsection{Lagrangian versus Eulerian descriptions \label{sec4-1}}

One can utilize either a Lagrangian or Eulerian description in the
analysis of a galaxy.  The Lagrangian description $\Gamma$ is,
for our purposes, defined to be the list of particles which
constitute a given galaxy.  $\Gamma$ conserves particle number
(i.e. mass) for all time, and does not allow for flux of particles
into or out of the system.  The Eulerian description of a galaxy is
taken to be the list of particles lying within the best-fit
ellipsoidal volume $V$ (described in the next subsection)
enclosing the Lagrangian list $\Gamma$.  $V$ can be made to conserve
mass if one requires that the volume encloses the same number of
particles as $\Gamma$; as a galaxy evolves and deforms, $V$ will vary 
significantly over time.  

\subsection{Principle axes and ellipsoidal fits \label{sec4-2}}

\begin{table}
\begin{center}
\caption{Statistics for elliptical axis fitting}
\begin{tabular}{|c|c|c|c|c|c|c|}
 Axis & & & \multicolumn{2}{c}{DC Method}
     &  \multicolumn{2}{c}{Our Method} \\
     Ratio & N & $\eta$
     & $\langle$  \% error $\rangle$
     & $\chi^{2}$ &  $\langle$  \% error $\rangle$
     & $\chi^{2}$ \\ \hline
$\ge0.1$ & 250&  0  &  $11. \pm 30.$ & 206.  & $4.4 \pm 3.5 $& 6.3  \\
       &      & $-2$&  $11. \pm 21.$ & 114.  & $5.3 \pm 4.3 $& 9.2  \\
       & 1000 &  0  &  $8.2 \pm 26.$ & 148.  & $2.2 \pm 2.2 $& 1.9  \\
       &      & $-2$&  $8.5 \pm 25.$ & 143.  & $2.8 \pm 2.4 $& 2.7  \\
       & 5000 &  0  &  $4.6 \pm 17.$ & 64.  & $1.1 \pm 1.8 $& 0.88 \\
       &      & $-2$&  $5.2 \pm 19.$ & 75.  & $1.4 \pm 1.7 $& 0.97 \\
$\ge0.8$ & 250&  0  &  $7.5 \pm 5.4$ & 8.6  & $4.6 \pm 3.3 $& 3.2  \\
       &      & $-2$&  $7.1 \pm 5.7$ & 8.2  & $5.6 \pm 3.9 $& 4.7  \\
       & 1000 &  0  &  $3.7 \pm 2.8$ & 2.2  & $2.2 \pm 1.6 $& 0.76 \\
       &      & $-2$&  $3.8 \pm 2.9$ & 2.3  & $2.6 \pm 2.0 $& 1.1  \\
       & 5000 &  0  &  $1.8 \pm 1.9$ & 0.68 & $0.96\pm 0.75$& 0.15 \\
       &      & $-2$&  $2.2 \pm 2.4$ & 1.1  & $1.1 \pm 0.90$&  0.22 \\
\hline
\end{tabular}
\label{tbl4-1}
\end{center}
\end{table}

The volume $V$ of a galaxy is determined from its ellipsoidal
parameters.   However, the accurate determination of axial ratios or
orientation angles of a 
discrete distribution of particles is far from trivial.  Many groups
(e.g. Splinter et al. 1997; de Theije, Katgert \& van Kampen 1995;
Plionis, Barrow \& Frenk 1991) 
use the fact that, for a uniform ellipsoid, the ratio of the
eigenvalues of the (normalized) inertia tensor is the square of the
ratio of the axes.  One can then simply fit spherical shells to halos
and determine the axis ratios from the normalized inertia tensor, as
in Frenk \etal (1988).   This method systematically
underestimates the axis ratios for nonspherical distributions.   Katz
(1991)  computes the axis ratios from the diagonalized inertia tensor
by iteratively deforming an initial spherical shell into an
ellipsoid. The axis ratios in each step are determined from the
particles contained within the ellipsoid of the previous step, however
the major axis is held at a fixed value for all steps.   Dubinski \& 
Carlberg (1991) (the DC method) iteratively perform a similar analysis 
in which the normalization parameter for the inertia tensor and the
axis ratios are recursively determined from each other.  Of all these
methods, the latter is the most aesthetically appealing since it
relies on the least number of assumptions (namely, only that the
density distribution is stratified in similar ellipsoids).  In test
scenarios using a $\rho \propto r^{-2}$ density profile, DC
find random errors between $1-10 $ percent, and note that there is a bias
toward underestimating the axis ratios if the intrinsic values are
$\ga 0.8$.   

To avoid appealing to initially spherical symmetries, we perform shape
analyses on the full bound-particle list $\Gamma$ for each galaxy. 
We determine the principle axes and Euler angles (used to rotate
between the body and Cartesian coordinate systems) by finding the
eigenvectors of the full three-dimensional inertia tensor,
\begin{equation}
 I_{ij}=\sum_{\alpha=1}^{M(\Gamma)} m_{\alpha} \left[ \delta_{ij}
        \sum_{k} x^{2}_{k}-x_{i}^{\alpha} x_{j}^{\alpha} \right].
	\label{eqn4-4}
\end{equation}
 Axes vectors 
$({\bmath a}, {\bmath b}, {\bmath c})$
are assigned such that $a \ge b \ge c$ and the Euler angles such that
the body ${\bmath a}$ axis is rotated onto the Cartesian ${\bmath x}$
axis.  Once the galaxy is translated and rotated onto the
Cartesian basis, the axis ratios are determined using the eigenvalues
$( e=\sqrt{\lambda_{2}/\lambda_{1}})$ of the the
two-dimensional inertia 
tensors for the particles projected onto the $x-y$ and $x-z$ planes
(effectively equivalent to the non-iterative method of Katz 1991).  
One has the option to iterate between the rotation and projected axis
ratios until a desired residual is achieved.  
The best-fit volume $V$ is now defined to be the ellipsoid using these
axis ratios and which contains the same mass as in the Lagrangian
volume $\Gamma$.   

We compared this method to the DC method for
test galaxies with random orientations and axis ratios, containing
between $N=$100 and 10000 particles and following density profiles $\rho
\propto r^{\eta}$ for $\eta \in [0,-3]$.  For each combination of
particle number and density power law, we ran one test with 1000
galaxies having random axis ratios $\ge0.1$ and a second test with 500
spheroidal galaxies having axis 
ratios $\ge 0.8$.  A subsample of the results are listed in Table
\ref{tbl4-1}, which shows the average error in the computed
axis ratios and $\chi^{2}$ (the sum of the squares of the percent
errors for all the fits).   In roughly every case except $\eta=-3$ (for
which both methods fail), our fitting routine achieves at least twice
the accuracy of the DC method, with a substantially lower scatter
about the mean.  We find that for the spheroidal galaxies, there is a
slightly higher tendency for both methods to underestimate (by $\la
10 $ percent) the actual axial ratios.  
We note that
this technique does not resolve radial structure, as in Warren et
al. (1992).  However this is also not our goal, given that the majority
of our galaxies contain less than 1000 particles per species.  

\subsection{Overdensity \label{sec4-3}}

For any given galaxy $n$ and epoch $t$, the overdensity 
$\delta^{(n)}(t)$ is calculated using the density $\rho^{(n)}(t)$ as
given by the best-fit ellipsoidal volume $V$ and the total mass of
particles within that volume.  At the earliest times, the
overdensities are extremely small and the calculation is not strictly
robust, mostly since identification of the volume exactly enclosing a
stochastic distribution is non-trivial.  At the initial epoch, we also 
calculate the initial overdensity threshold parameter $\nu$, defined
such that $\delta_{i}(M) = \nu \sigma(M)$, where $\sigma(M)$ is the
standard deviation of the overdensity as measured on the mass scale of
the galaxy in question.  For each galaxy of mass $M$, we determine
$\sigma(M)$ by sampling the overdensity of $10,\!000$ randomly placed
spheres, each also containing $M$ particles.  
This differs from BE87, as Hoffman (1988) has pointed out that their
method may introduce bias in $\nu$ by calculating $\sigma$ on a
physical smoothing scale rather than a galaxy's mass scale.

\subsection{Predicted and actual angular momentum \label{sec4-4}}

We compute ${\bmath L}(t)$ directly for each galaxy from its
Lagrangian particle list $\Gamma$ (containing $M$ particles) using
\begin{equation}
 {\bmath L}=\sum_{\alpha=1}^{M}({\bmath r}_{\alpha}
 \times m_{\alpha}{\bmath v}_{\alpha}).  \label{eqn4-1}
\end{equation}
We calculate the predicted angular momentum from equation
(\ref{eqn2-5}) by computing the deformation tensor ${\cal D}$ and
inertia tensor ${\cal I}$ 
for each galaxy at the initial epoch {\em only}, and evolve their
tensor product forward using the time-dependent factor $a^{2}\dot{D}$.
The deformation tensor, as given by equation (\ref{eqn2-8}), requires
the Fourier transform of the potential 
$\tilde{\psi}({\bmath k})$.
The entire mass distribution is first smoothed onto a uniform grid
using the three-dimensional S2 window function (Hockney \& Eastwood,
1981), 
\begin{equation}
  W(r,a)=\left\{ \begin{array}{ll}
	\frac{48}{\pi a^{4}}\left(\frac{a}{2}-r\right), & \mbox{if $r <
	\frac{a}{2}$}, \\ 
	0, & \mbox{otherwise}.
  \end{array} \right.
  \label{eqn4-2}
\end{equation}
Note that our S2 smoothing function utilizes the full 3-D smoothing of a
spherical volume onto a grid, as opposed to using three 1-D smoothings
as is done, for example, in the triangular shaped clouds method.

The resulting density field is Fourier transformed and convolved with
a Greens' function to yield the $k$-space potential, which is
evaluated in equation (\ref{eqn2-8}) along with the Fourier transform of
the S2 function
\begin{equation}
 \tilde{W}(k,a)=\frac{12}{(ka/2)^{4}}\left(2-2\cos{\frac{ka}{2}}
 -\frac{ka}{2}\sin{\frac{ka}{2}}\right). \label{eqn4-3}
\end{equation}
Values for $L$ and $L^{(1)}$ will be given in the
unconventional units of $10^{12} \msun$ kpc$^{2}$ Gyr$^{-1}$.

\subsection{Turnaround and crossing times \label{sec4-5}}

Determination of the turnaround time is another nontrivial problem.
Indeed, it is unclear whether an evolving protogalaxy ever passes
through a period of maximal isotropic expansion, or whether all
material at a given distance from the centre undergoes turnaround at a
common time (Peebles, 1980, p.\ 86).  Any prediction (\eg the
final angular momentum) that is based upon an ill-defined dynamical
point during evolution (\eg the angular momentum at turnaround) will
necessarily be skewed by the uncertainty in determining that epoch.
It is therefore important to understand both the uncertainty involved
in localizing the turnaround epoch and how this propagates through
linear predictions.

To explore statistical consistency, we define four
independent methods to determine the turnaround time for
each galaxy: the first two empirical, the third semi-empirical, and the
latter analytic.  In the first method, we calculate the average radial
velocity of particles with respect to the galaxy's centre of mass.
Turnaround $t_{Md}$
is defined as the time when this velocity divergence inverts from
positive (expansion) to negative (contraction).  The second method
defines turnaround $t_{M\%}$ as the earliest time at which at
least half the particles in a galaxy are infalling.  The
semi-empirical method measures the overdensity of the
galaxy at each epoch and defines the turnaround $t_{M\delta}$ as the time
at which $\delta(t) \simeq 4.55$ (from the spherical-collapse model,
\S2).  The analytic method measures the 
initial overdensity of each galaxy and calculates the epoch $t_{Mz}$
from equation (\ref{eqn2-12}) such that 
\begin{equation}
  z_{M}=[(20/3)\delta_{i}(1+z_{i})(6 \pi)^{-2/3}] - 1, \label{eqn4-5}
\end{equation}
where subscript $i$ indicates the initial value.  

Similarly, we use four methods to determine the time of caustic
crossing.  As explained above, our first-order definition is 
$t_{Cto} = 2 \langle t_{M} \rangle$.  Once virialized, the galaxy will
have shrunk by a factor of 2, thereby increasing its density
eightfold, and we expect (again to first order) that $\delta_{C}
\simeq 8\delta_{M} = 43$ at $t_{C\delta}$.  When computing the
velocity divergence, we also calculate the variance in the peculiar
velocities of all particles in a galaxy, and we expect that this value
will be highest (i.e. the most disorder in the velocity field) when
the galaxy experiences 
shell crossing ($t_{C\sigma}$).  Finally, if an ellipsoidal
proto-galaxy is rotating then we expect its initial collapse to
occur first along the short axis followed by the intermediate and long
axes, and resulting first in a planar or pancake
structure (Zel'dovich 1970; Peebles 1980 \S 20) followed by a prolate
ellipsoid.  We define 
$t_{Cax}$ as the epoch at which the major axis $a$ is longest compared
to the other two, which should indicate the shell crossings of the
shorter axes.  For both turnaround
and shell crossing, we use the median of the four methods as the
final value ($\langle t_{M} \rangle$,$\langle t_{C} \rangle$
respectively), since the median minimizes the effects of single
outliers. 

\section{Results \label{sec5}}

In SPH simulations, centrally-located gas particles in circular motion
experience an artificial pressure due to the finite smoothing
length of the hydrodynamic forces, and therefore tend to clump within
the inner $\sim1.5$ smoothing radii from the centre, which typically
represents the size of a galactic disk in this simulation.  
As such, in this study gas particles are used only to identify the most
probable sites of galaxy formation, while dark matter dominates the
overall dynamics and is therefore used exclusively in this
analysis.  Dynamical effects of the gas component on dark halo
evolution will be addressed in paper II.
Furthermore, it is possible that the lowest-mass galaxies ($M \la
200$ DM particles) contain dynamical masses too small for stable numerical
resolution.  When applicable, we divide the sample of galaxies into
subsets with $M>200$ or $M>1000$ DM particles, with the understanding
that better-resolved galaxies should naturally provide more consistent
results. 

\subsection{Collapse history \label{sec5-1}}

\begin{figure*}
\epsfig{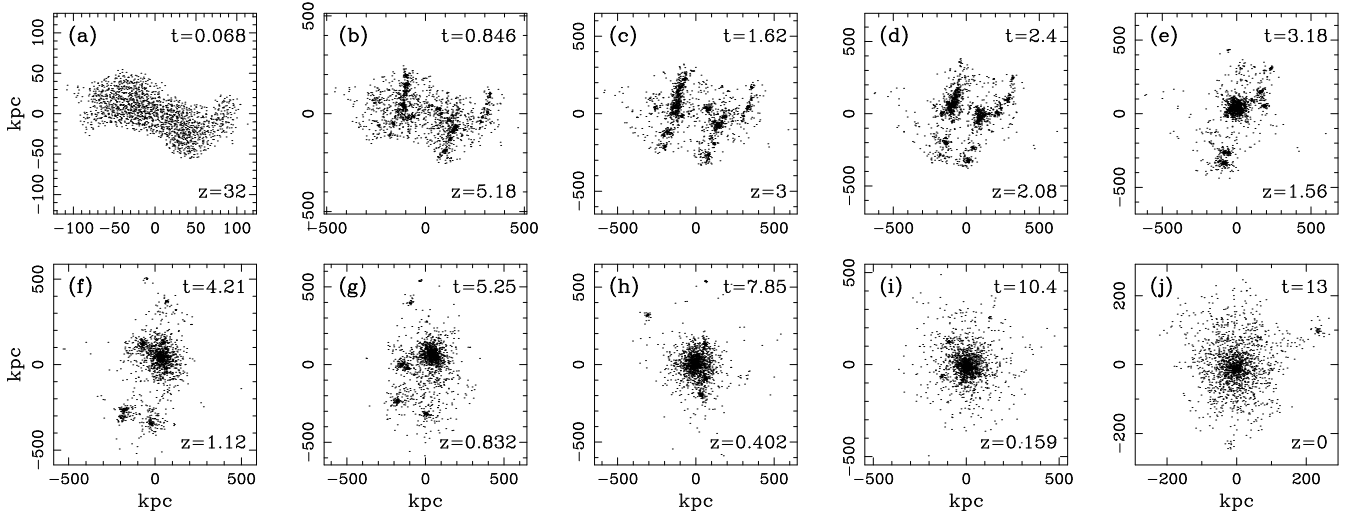}
\caption[gal12]{Particle positions plotted in physical coordinates 
 and projected onto the Cartesian
 $(\hat{\imath},\hat{k})$ plane for galaxy 12.  The epoch is listed in
 Gyr in the upper right corner and the redshift in the lower right
 corner of each panel.  Turnaround and crossing times correspond
 roughly to panels [c] and [e], respectively.}
\label{gal12}
\end{figure*}

\begin{figure*}
\epsfig{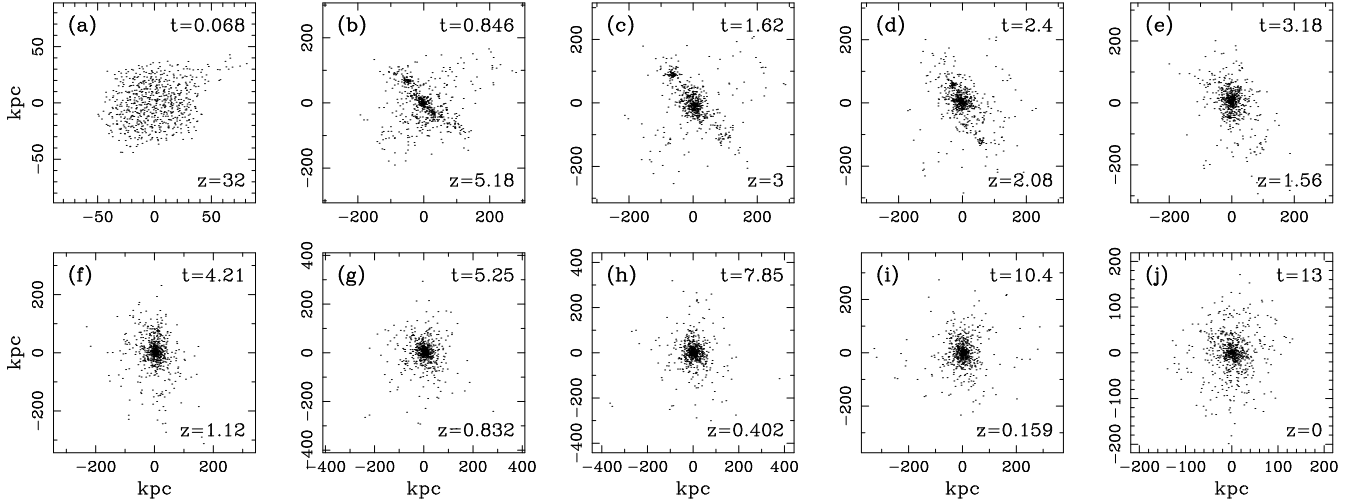}
\caption[gal22]{Same as Fig.\ \ref{gal12} for galaxy 22.  Turnaround and
 crossing times correspond roughly to panels [b] and [d] respectively.}
\label{gal22}
\end{figure*}

We first study the spatial evolution of each galaxy in our catalog by
following the particle list $\Gamma$ from the initial to the final
epoch.  Figures \ref{gal12} and \ref{gal22} show the evolution of 
galaxies \#12 and \#22, containing 1776 and 703 DM particles
(respectively).  It is
evident from these two evolutionary sequences that the collapse is
neither spherical nor homogeneous.  Rather, the observed evolution,
typical of all the galaxies in our catalog,  follows that predicted
from the hierarchical-clustering scenario in the CDM model, in which
a relatively isotropic distribution first collapses into small
objects, which then subsequently accrete and virialize into the final
bound galaxy.   

\subsubsection{Turnaround \label{sec5-1-1}}

\begin{table}
\caption{Sample turnaround and caustic crossing times}
\begin{tabular}{c l c c c c c}
 &  
     & \multicolumn{1}{c}{$t_{Md}$} 
     & \multicolumn{1}{c}{$t_{M\delta}$} 
     & \multicolumn{1}{c}{$t_{Mz}$} 
     & \multicolumn{1}{c}{$t_{M\%}$}  
     & \multicolumn{1}{c}{$\langle t_{M} \rangle$} \\
{Galaxy} & {Mass} & {$t$[Gyr]}  
       & {$t$[Gyr]} 
       & {$t$[Gyr]} 
       & {$t$[Gyr]} 
       & {$t$[Gyr]} \\ \hline
  8 &  2431 &   2.92 &   2.66 &   2.92 &   2.92 &   2.92 \\ 
 12 &  1776 &   1.56 &   2.14 &   1.62 &   1.56 &   1.62 \\ 
 22 &   703 &   0.91 &   1.24 &   1.24 &   0.91 &   1.11 \\ 
 30 &   441 &   0.85 &   1.24 &   0.98 &   0.78 &   0.98 \\ 
 32 &   424 &   1.82 &   1.95 &   1.49 &   1.82 &   1.82 \\ 
\hline \hline
    &  
     & \multicolumn{1}{c}{$t_{Cto}$} 
     & \multicolumn{1}{c}{$t_{C\delta}$} 
     & \multicolumn{1}{c}{$t_{Cax}$} 
     & \multicolumn{1}{c}{$t_{C\sigma}$} 
     & \multicolumn{1}{c}{$\langle t_{C} \rangle$} \\
   {Galaxy} & {Mass} & {$t$[Gyr]} & {$t$[Gyr]} & {$t$[Gyr]} & {$t$[Gyr]} 
       & {$t$[Gyr]} \\ \hline 
  8 &  2431 &   5.77 &   6.55 &   2.66 &   4.73 &   4.73 \\ 
 12 &  1776 &   3.18 &   4.99 &   3.44 &   3.44 &   3.18 \\ 
 22 &   703 &   2.14 &   3.18 &   3.18 &   2.66 &   2.60 \\ 
 30 &   441 &   1.88 &   4.48 &   2.47 &   1.75 &   2.34 \\ 
 32 &   424 &   3.44 &   4.48 &   4.48 &   3.44 &   3.70 \\ 
\hline
\end{tabular}
\label{tbl5-1}
\end{table}

As a predictive tool, the linear theory [equation (\ref{eqn2-5})]
depends upon a reasonable determination of the turnaround time.
Of the four methods listed in \S \ref{sec4-5}, the analytic method
$t_{Mz}$ predicts the turnaround epoch from only the initial
conditions, while the others use measured quantities from the evolving
simulation.  In view of the facts that the initial particle
distributions are non-spherical, that collapse is hierarchical, and
that turnaround for an ellipsoidal distribution is ill-defined, it is
essential to address whether the spherical-collapse model can
provide any accurate determination of the turnaround time of a
galaxy.  

We present the computed turnaround and caustic crossing times
for a variety of distinctly different evolutionary histories in Table
\ref{tbl5-1}.
Galaxies \#12 and \#22 (well-behaved evolution \S
\ref{sec5-2}) are complimented with a large
merger (\#8) and galaxies \#30 and \#32, which suffer strong
tidal encounters (\S\ref{sec5-2-2}).
The respective four methods for determining each of
these evolutionary epochs (described in \S \ref{sec4-5}) are all
relatively consistent for the galaxies listed, independent of the
large variation in collapse history. 

\begin{table*}
\begin{minipage}{100mm}
\caption{Statistics for $t_{M}$ and $t_{C}$ }
\label{tbl5-3}
\begin{tabular}{c c c c c c c}
   {Catalog} & {$\overline{\sigma(t_{M})}$} & 
   {$\overline{\sigma(t_{C})}$} & 
   {$\overline{\langle t_{M} \rangle}$} & 
   {$\overline{\langle t_{C} \rangle}$} &
   {$\overline{\langle t_{M} \rangle/t_{Mz}}$} &
   {$\overline{\langle t_{C} \rangle/t_{Mz}}$} \\ \hline
All \fullcat 
       & $0.38$ & $1.39$ & $1.54 \pm 0.61$
       & $ 3.07 \pm 1.07$ & $1.13 \pm 0.41$ & $2.27 \pm 0.71$ \\
LG \cutcat    
       & $0.35$ & $1.34$ & $1.51 \pm 0.61$
       & $3.00 \pm 1.05$ & $1.07 \pm 0.33$ & $2.17 \pm 0.61$ \\
SE \stdcat   
       & $0.34$ & $1.38$ & $1.48 \pm 0.59$ 
       & $3.02 \pm 1.09$ & $1.02 \pm 0.32$ & $2.10 \pm 0.64$ \\
\hline
\end{tabular}

\medskip
{Columns 2 \& 3 give the average dispersion between the
four epochs for determining $t_{M}$ and $t_{C}$.   Columns 4 \& 5
list the average over each catalog of the median values of the
previous four epochs.  Columns 6 \& 7 list the average over each
catalog of the median epoch divided by the analytically-determined
$t_{Mz}$.  See \S\ref{sec4-5}} 
\end{minipage}
\end{table*}

Statistics for the full galaxy catalog, as well as statistics for
subsamples of \cutcat and \stdcat
galaxies with well-behaved evolution (\S \ref{sec5-2}) are
listed in Table \ref{tbl5-3}.  The first pair of data columns show
that the mean dispersion between the epochs given by the four methods 
is relatively small ($\sim 0.4$ Gyr) for turnaround and slightly larger
($\sim 1.4$ Gyr) for caustic crossing.  Recall from
\S \ref{sec4-5} that the final value $\langle t_{M} \rangle$ of
turnaround (caustic-crossing) is the median of the four methods;
the mean of these median values is given in the second pair of
columns.  A priori, we do not expect the turnaround (caustic
crossing) epochs to 
be the same for all galaxies, hence the large dispersion in these
values.  The average caustic crossing time is nearly double that for
turnaround, reflecting both our input assumption (\S2) and
that the spherical-collapse model is consistent with early
post-turnaround evolution. 

The last pair of columns in Table \ref{tbl5-3} give
the average ratio of the median turnaround (caustic crossing) time to
the analytic turnaround epoch $t_{Mz}$.  That the average value of 
\mbox{$\langle t_{M} \rangle / t_{Mz} \sim 1$} with very little
dispersion shows that $t_{Mz}$ is highly consistent with the empirical
measures of turnaround; if we believe these latter three values,
then the spherical-collapse model successfully predicts the
turnaround epoch.  Further, we find that the average caustic
crossing time is nearly double $t_{Mz}$, also as expected.  These
tests support the applicability of the spherical-collapse model for
determination of the turnaround and caustic crossing times.   

In both Figures \ref{gal12} and \ref{gal22}, one clearly observes
that the particles pass through a planar or Zel'dovich (1970) pancake
stage at extremely early times.  Kuhlman, Melott \& Shandarin (1996)
find that the early collapse of the short axis and growth of the long
axis are generic trends of particle distributions.  
EL95 also find that the short axis
of a rotating ellipsoid typically collapses first into the expected
sheet-like structure.  This planar stage generally correlates closely with
the predicted turnaround for a galaxy.   Shell crossing occurs
dramatically in galaxy \#12 when the two pancakes collide, and more
subtly in galaxy \#22 as the intermediate axis collapses.  After the
collapse of the short axis, the galaxy still retains a significant
quadrupole moment, since the quadrupoles are proportional to the
difference between the squares of the axis lengths (Peebles 1969).
Considering that the acquisition of angular momentum
results from a coupling between the large-scale tidal field and the
quadrupole moment of the galaxy, this coupling will remain significant
until the longer axes have collapsed.  We therefore expect the angular
momentum in these non-spherical galaxies to increase beyond the
spherically-defined turnaround epoch.  In the linear
approximation [eq.\ (\ref{eqn2-6})], we use only the initial values of
the deformation and inertia tensors, having assumed that during
linear evolution, the timescales over which these tensors change are
long compared to that of turnaround.  As the axes collapse, the
inertia tensor and hence the quadrupole moment will change strongly
from its initial value, and it is therefore unclear whether we expect
the growth of angular momentum near or beyond turnaround to continue
linearly.   

\subsubsection{Shape analysis \label{sec5-1-2}}

\begin{figure*} 
\epsfig{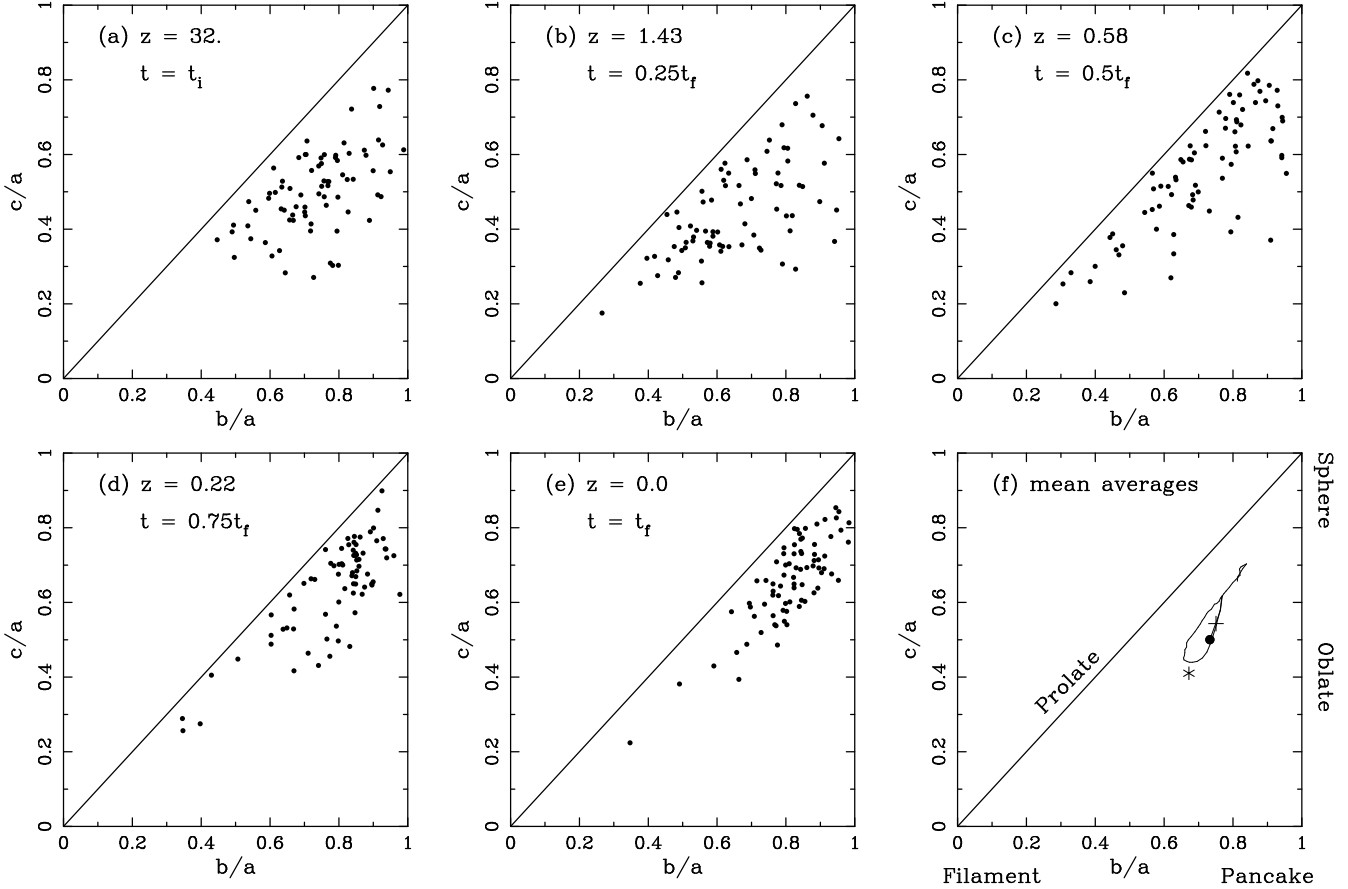}
 \caption{Distibutions of axis ratios for the full catalog of
 galaxies at the (a) initial, (b)-(d) intermediate, and (e)
 final epochs.  (f) The mean axis ratio for the full catalog of
 galaxies is plotted for all epochs, starting at the solid dot.  After
 a brief rise (motion with the Hubble expansion), the average galaxy
 collapses toward a pancake structure near turnaround (cross mark),
 then toward a prolate ellipsoid during shell crossing (star mark) and
 relaxes into a triaxial system.  This Figure may be compared with
 Eisenstein \& Loeb (1995) Figure 2. }
\label{axrat}
\end{figure*}

To aid in studying the shape evolution of our galaxies, we define the
morphological triaxiality parameter as given by Franx, Illingworth \&
de Zeeuw (1991), 
\begin{equation}
 T=\frac{a^{2}-b^{2}}{a^{2}-c^{2}}, \label{eqn5-1}
\end{equation}
for which $T=1$ is purely prolate and $T=0$ is purely oblate.  One may
then classify galaxies with $0<T<\frac{1}{3}$ as oblate,
$\frac{1}{3} < T < \frac{2}{3}$ as triaxial, and $\frac{2}{3}<T<1$ as
prolate ellipsoids.  

For initial conditions similar to ours, EL95
find their initial halos to be prolate-triaxial.  
Given that the applicable collapse model is, to a better
approximation, a collapsing ellipsoid rather than a sphere, we expect
the following generic evolution of a galaxy: the initial
prolate-triaxial particle distribution first expands linearly with the
Hubble flow, and possibly becomes slightly more spheroidal.  As the
short axis collapses, the galaxy moves toward the oblate-pancake
state, until the secondary axis collapses, at which point 
the galaxy evolves toward a prolate ellipsoid.  At caustic crossing,
the galaxy should pass through its most prolate distribution, after which
virialization processes relax the system into an eventually spheroidal
state.  Unless a galaxy has suffered an extremely large merging event,
in which case it is generally still relaxing or merging at the final 
epoch, our galaxies have generally accreted the majority of their
clumps by roughly $t=\frac{1}{2} t_{f}$.  Therefore we expect the
galaxies to relax and virialize during the second half of the
simulation. 

We plot the evolution of the axis ratios in Figure \ref{axrat} for all
\fullcat galaxies in our catalog.  This may be compared
with Figure 2 of EL95 except they evolve their
galaxies only until the caustic crossing time.  We see the expected
initial quasi-spherical state (panel [a]) which evolves toward a
prolate ellipsoid (panel [b]) and returns toward a spheroid
(panels [c]-[e]).  The mean values for the galaxies at each output are
plotted in panel [f].  Each galaxy evolves at its own rate, therefore
this latter panel is not a completely accurate representation of the
axial history.  Nonetheless, we see the expected pattern from above.
From the initial epoch, axial ratios briefly move toward the
spheroidal state before collapsing toward first a pancake and then a
prolate ellipsoid, after which they return toward a spheroidal
distribution.   

\begin{figure} 
\epsfig{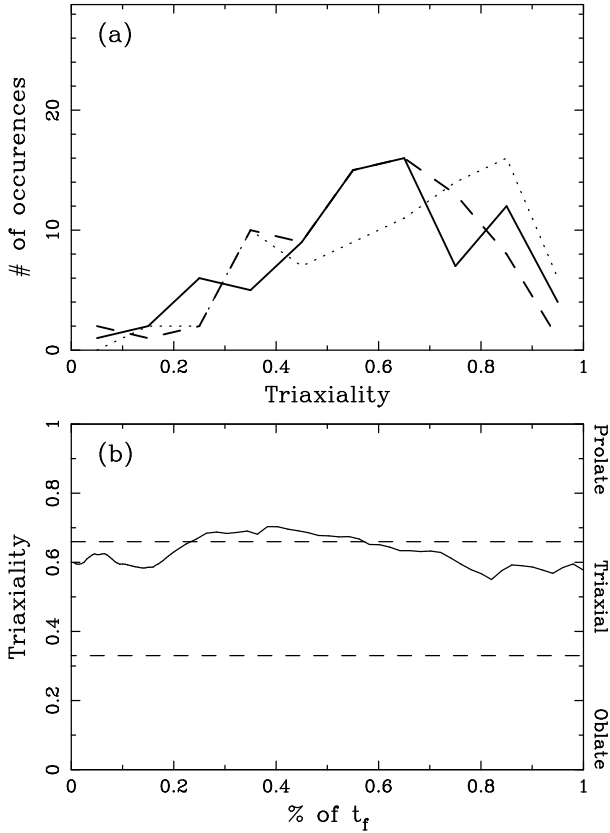}
 \caption{(a) Binned counts of triaxiality for the full galaxy catalog
 at different epochs: $t_{i}$ (solid curve); average epoch of caustic
 crossing (($t=2.68$ Gyr; dotted curve); $t_{f}$ (dotted curve).  (b)
 Mean triaxiality for the full 
 galaxy catalog plotted versus epoch.  The distribution is 
 predominantly prolate at $1/4t_{f}$ and triaxial at $t_{f}$
 See \S\ref{sec5-1-2}. }
\label{triax}
\end{figure}

The evolution of the triaxiality parameter is presented in Figure
\ref{triax}.  Histograms of the triaxiality parameters are given in
panel [a] for $t_{i}$, the average epoch of caustic
crossing ($\langle t_{C} \rangle =2.68$ Gyr) and $t_{f}$.
The time evolution of the average values of the triaxiality parameters
are given in panel [b].  As expected, the initial galaxies are
prolate-triaxial, and evolve to a more prolate state at
caustic crossing.  The final state of the galaxies is predominantly
triaxial with a wide tail of both prolate and oblate ellipsoids. 
For comparison, Warren \etal (1992) find that their halos
are predominantly prolate.  Dubinski (1994) finds that a coupling
between the orbital distribution of dark matter and dissipative infall
of gas (post shell-crossing) generally transforms a
prolate-triaxial halo  
$(T \sim 0.8)$ to an oblate-triaxial halo $(T \sim 0.5)$, which we
observe in Figure \ref{triax}[b].  Dubinski further notes that
the oblateness of the dark halo morphology constrains evolved
halos to have $b/a \ga 0.7$, which we see in Figure
\ref{axrat}[f].  Finally he finds that the flattening $c/a$
increases from roughly 0.4 to 0.6, which we also measure post
shell-crossing, and which 
further agrees with deprojection analyses of observed elliptical
galaxies $\langle c/a \rangle \sim 0.65-0.7$ (Binney \& de Vaucouleur
1981; Frank \etal 1991; Ryden 1992; Frasano \& Vio 1991).

Lastly, we find that the final shape of the galaxy
has little or no dependence on the initial distribution, and therefore
there does not appear to exist a bijective mapping between initial and
final conditions.  Furthermore, this implies that the mapping between 
Eulerian and Lagrangian descriptions of galaxies are unique
to each particular galaxy and epoch. 

\subsection{Evolution of angular momentum \label{sec5-2}}

Sample plots of the total angular momentum $(L=|{\bmath L}|)$ evolution
are shown in Figures \ref{Loft1} and \ref{Loft2}.  The five galaxies
discussed in the 
previous section (\#s 12, 22, 8, 30, 32) are displayed in panels
\ref{Loft1}[a]-[b] and \ref{Loft2}[a]-[c] respectively.  We have 
marked the four turnaround epochs and caustic crossing epochs (see
Figure caption for legend) on each curve.  In galaxy \#12, each 
quadruplet of points for both evolutionary epochs are nearly
coincident.  Despite the wide range of evolutionary histories in the
remaining plots, the dispersion among these sets of points is quite
small for each galaxy.  This result supports the
conclusions from \S \ref{sec5-1-1} that our
determination of these epochs is self-consistent and that the 
spherical-collapse model can quantitatively describe these highly
non-spherical processes, including a wide mass range and interacting
systems. 

\subsubsection{Qualitative comparison with linear theory \label{sec5-2-1}}

Linear theory coupled with the spherical-collapse model predicts that
the angular momentum of an evolving galaxy 
will grow linearly until turnaround.  From our discussion
in \S \ref{sec5-1-1} we expect the angular momentum to grow past
turnaround and potentially until caustic crossing.  Barring merging
and accretion events, the angular momentum of a bound galaxy should
remain constant in time thereafter.

\begin{figure*} 
\epsfig{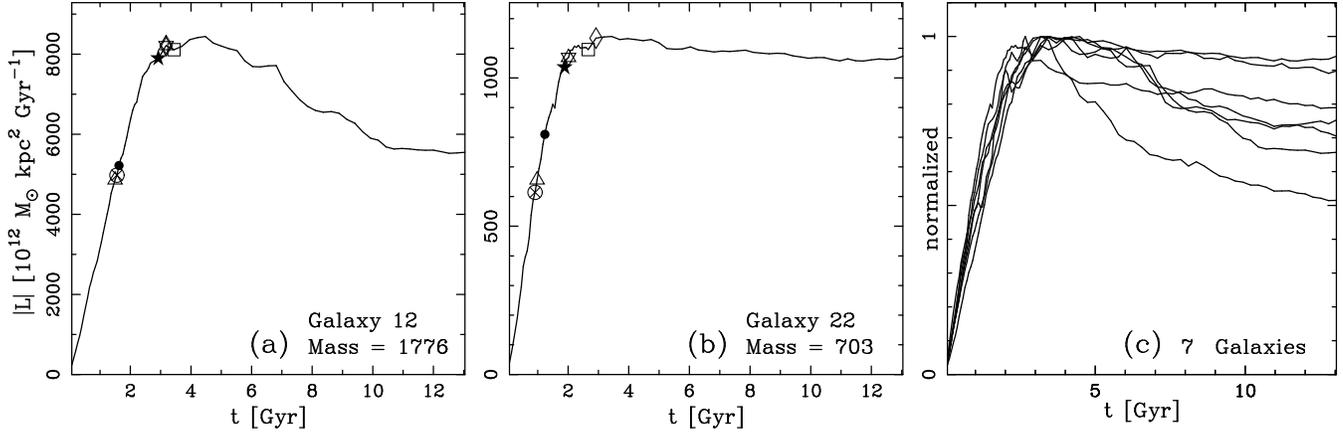}
 \caption{Total angular momentum versus epoch.  (a)-(b) Examples of
 the ``standard evolution'' of angular momentum, also marked with the
 measured turnaround and shell crossing epochs.  (c) Overlay of seven
 normalized standard evolution galaxies.  Note the general features of
 linear growth beyond turnaround, and decay of angular momentum at
 later times.  Small perturbations in the evolution correspond to
 minor tidal encounters with neighbors.  
 Markers: cross = $t_{Md}$; triangle = $t_{M\delta}$; solid
 dot = $t_{Mz}$; open circle =  $t_{M\%}$; solid star = $t_{C\sigma}$;
 diamond = $t_{Cax}$; open star = $t_{Cto}$; square = $t_{C\delta}$. }
 \label{Loft1}
\end{figure*}

\begin{figure*} 
 \epsfig{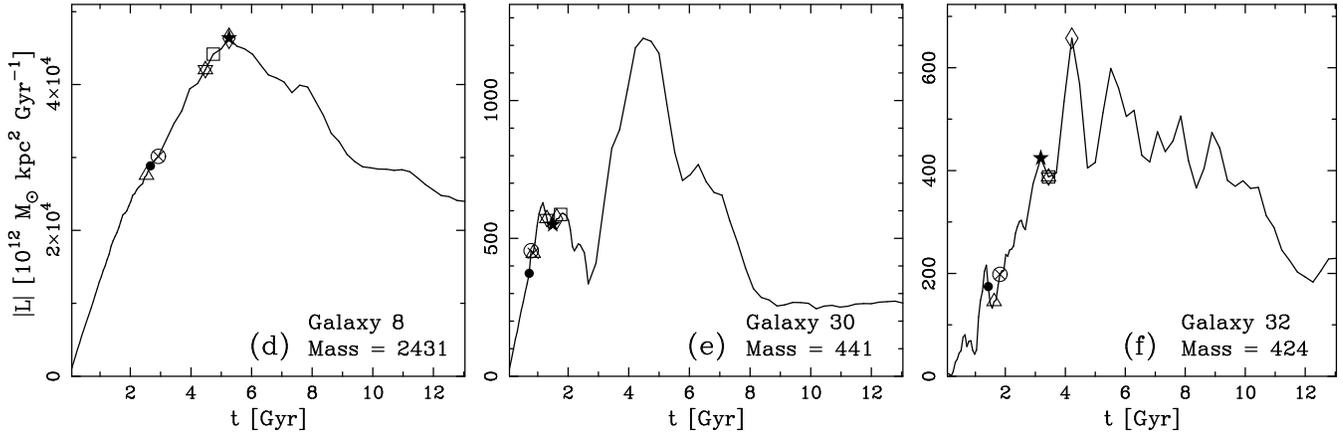}
 \caption{Notation as in Fig. \ref{Loft2}. (a) A heavily merging system.
 (b)-(c) Galaxies suffering large-scale and repeated tidal
 encounters.  }
 \label{Loft2}
\end{figure*}

We find that the expected model just described is only partially
followed by the generic observed evolution of angular momentum.
The normalized curves of $L(t)$ for 7 galaxies have been overlayed in
Figure \ref{Loft1}[c], and describe what we term the 
``standard evolution.''  $L(t)$ first rises linearly, up to and often
beyond the turnaround epoch.  The maximum value is reached near
caustic crossing, after which the angular momentum turns over and
slowly decays with time.  \cutcat of our galaxies exhibit smooth
linear growth of angular momentum before turnaround, and of these,
\stdcat evolve according to the standard evolutionary sequence
described in panel [c] (the remaining galaxies will be addressed
shortly).  We identify the \cutcat galaxies as the ``LG'' catalog (for
linear growth) and the subset of \stdcat as the ``SE'' catalog (for
standard evolution), while the \exccat remaining galaxies are grouped
in the ``INT'' (for interacting, \S \ref{sec5-2-2}) catalog.  

The average time-dependence of $L$ for the full, LG, SE and INT
catalogs are given in
Table \ref{tbl5-4}.  The pre-turnaround evolution of $L$ scales as $t$
to within 5 percent.  $L(t)$ deviates more strongly from
linearity as it begins to turn over after $t_{M}$.  On average, 
$L(t_{C}) \sim (2-3)L(t_{M})$ and since $t_{C} \cong 2t_{M}$, the
growth of $L$ is still approximately linear [$L(t<t_{C}) \propto
t^{0.85}$] between turnaround and caustic crossing.  Given the
possible non-spherical processes that could cause $L(t)$ to grow
non-linearly at early times (examined in \S \ref{sec5-1-2}), it is
doubly significant that the observed evolution of angular momentum so
closely matches that predicted from linear theory, since this
indicates that the considerable non-spherical processes have minimal
effect during linear growth.

\begin{table}
\caption{Average $\langle \beta \rangle$ in $L \propto t^{\beta}$}
\label{tbl5-4}
\begin{tabular}{c r r r}
  {Catalog} & {$\beta$ until $t_{M}$} & 
  { $\frac{1}{2} \left( t_{M}+t_{C} \right)$} &
  {$\beta$ until $t_{C}$}
\\ \hline
All \fullcat 
   	& $0.93 \pm 0.14$ & $0.87 \pm 0.16$ & $0.83 \pm 0.16$ \\
LG \cutcat    
	& $0.95 \pm 0.10$ & $0.90 \pm 0.13$ & $0.84 \pm 0.14$ \\
SE \stdcat   
	& $0.95 \pm 0.09$ & $0.90 \pm 0.13$ & $0.84 \pm 0.14$ \\
INT \exccat 
	& $0.90 \pm 0.18$ & $0.83 \pm 0.21$ & $0.79 \pm 0.23$ \\
\hline
\end{tabular}
\end{table}

Simple physics dictates that the angular momentum of an isolated
system must remain constant, and therefore if a 
post-turnaround galaxy is indeed largely insensitive to external tidal
fields (Peebles 1980) and relatively isolated,
we expect $L(t)$ to remain roughly constant for $t \ga
t_{M}$ for spherical collapse, and $t \ga t_{C}$ for 
ellipsoidal collapse.  Rather, we find the general trend, as seen in
Figure \ref{Loft1} (and in nearly every galaxy in our full catalog),
that the measured angular momentum decays after shell crossing.  
BE87 also find this trend in their simulations, and attribute it to
the loss of angular momentum to surrounding material during merging
and ``highly nonlinear evolution.''

We examined the mechanism for loss of angular momentum by studying
galaxies for which we had not culled out unbound particles after
running the FOF group-finding algorithm.  After $t_{C}$, the decay of
angular momentum is much more pronounced.  Identically, in galaxies for
which we exclude a subset of the outermost bound particles, the
angular momentum also decreases more strongly 
with time.  Since it is computationally too expensive to determine
{\em every} bound particle within a group (\S \ref{sec3-2}) it is expected
that we will not identify every particle bound to a given system using
our group-finding technique.  Each gravitationally-bound 
galaxy contains particles at radii greater than our effective
overdensity cutoff which have not been included in our catalog, 
and to which significant amounts of angular momentum are transferred
after shell crossing and during relaxation.  This transfer of angular
momentum from triaxial distributions to outlying particles is expected
through Landau damping in self-gravitating systems.  Unless
every particle bound to a group can be identified, the apparent decay
of the measured angular momentum will be a general trend of numerical 
simulations.  

\subsubsection{Effects of the local tidal field \label{sec5-2-2}} 

The question still remains why 43 of our \fullcat galaxies which pass
the original selection criteria do not have angular momenta matching 
the standard evolution model [due to noisy $L(t \ga t_{C})$], and
why \exccat of these have angular momenta which are not smoothly
linear at early times.  As an approximate study of the
evolution of the tidal field, we compute the torque on each galaxy by
every other galaxy in the simulation (a short $N^{2}$ calculation
since there are only 98 identified groups).  Substituting the first
three terms of the Taylor expansion of the potential into the torque
integral and simplifying, we find 
\begin{equation}
 \tau_{i} = \epsilon_{ijk} T_{jl} {\cal I}_{lk}, \label{eqn2-17}
\end{equation}
where ${\cal I}_{jk}$ is the inertial tensor (equation \ref{eqn2-7}) and 
(Dubinski 1992).\footnote{It follows that for a potential field
$\psi$, $T_{ij} = \left[ 3\partial_{i}\partial_{j}-\delta_{ij}\nabla^{2}
         \right]\psi$
from which we see directly that
the the deformation tensor [equation (\ref{eqn2-6})] is directly
related to the gravitational tidal tensor, since the diagonal elements
of both tensors do not contribute to the antisymmetric tensor product
with the inertial tensor (CT96).}
\begin{equation}
 T_{ij} = G \sum_{\alpha} m_{\alpha} \frac{3x_{i}^{\alpha}
  x_{j}^{\alpha}-\delta_{ij}r_{\alpha}^{2}}{r_{\alpha}^{5}}.
 \label{eqn2-14} 
\end{equation}
Since we are not measuring the torque due to the full mass
distribution in the simulation, we do not expect $\tau$ to remain
constant before turnaround, as would be expected from the linear
growth of $L$. Conversely, for a universe consisting of only two
galaxies with fixed comoving separation and no other external tidal
forces, we expect the torque to decay as $\tau \propto t^{-2}$. 

The torque acting on a typical galaxy in the SE catalog roughly
follows a $t^{-2}$ decay superposed with small perturbations 
($\la 50 $ percent) reflecting weak encounters with neighbors, generally
at large distances [$\ga (1-2)$ Mpc]. These typically correspond
to low-level ($\la 10 $ percent) fluctuations in the   
angular momentum, especially during late times ($t \ga t_{C}$).  
Intuitively, these are the expected conditions for relatively isolated
field galaxies.  Since the roles of the tidal field and induced shear
in the quasi- and non-linear regimes of these galaxies are not
negligible, models of single systems (\eg Zaroubi, Naim \& Hoffman
1996) should explore the quantitative effects of inclusion versus 
exclusion of tidal interactions.

Naturally, however, not all galaxies will exist in the
``field.''  Every galaxy excluded from our SE catalog suffers strong
(1-2 orders of magnitude) and repeated tidal interactions with 
neighboring galaxies.  The angular momenta of these galaxies can vary
by as much as $50 $ percent over $1-2$ Gyr periods throughout the entire
simulation.  These are 
galaxies expected to be found in groups or clusters, with the most
massive examples representing the central members.  Galaxy 30, shown in
Figure \ref{Loft2}[b], orbits within a few hundred kpc of a galaxy three
times as massive, while suffering minor encounters with comparably
sized galaxies at distances of 1-2 Mpc.  While the late-time evolution
is highly erratic, the early growth of $L$ is still linear.  Galaxy
32, shown in panel [c], suffers two close encounters with galaxies
three and five times as 
massive. Here, the rapid variations in $L$ appear quasi-periodic, most
likely reflecting the rotational frequency of the galaxy.\footnote{$L$
is expected to vary at roughly twice the rotational frequency of the 
torqued object.  This is also seen in the tidal interactions of binary
stars.}  Even in this extreme case, the growth of $L$ is still linear
at early times, although heavily polluted due to the effects of early
tidal interactions.  This is the case with all \exccat galaxies in the
INT catalog (hence the name ``interacting'').  Despite the noisy early
evolution, the average pre-turnaround angular momentum grows as
$t^{0.928}$ (see Table \ref{tbl5-4}) for these galaxies.  
Therefore, even in the rich tidal environment
of clusters, we find (quasi)linear growth of $L$ at early times,
indicating that the local tidal field is not negligible, but that the
effects do not become dominant until local evolution becomes
non-linear.   

We also expect that galaxies may suffer tidal interactions from
the outermost shells of their own infalling matter.  In galaxies
originating from higher-$\nu$ peaks (defined in \S \ref{sec4-3}), the
matter in  the immediately surrounding low-$\nu$ region will have a
density  similar to that of the collapsing object at early times.
According to the formalism of the 
spherical-collapse model, this corresponds to a mass shell of
$\Omega \ga 1$ but $<\Omega_{galaxy}$.  This shell will follow the
collapse of the galaxy and will evolve into surrounding matter
that is closer than its initial comoving distance,
generating more torque than expected from linear theory
(EL95).  Even our most isolated galaxies still
undergo small fluctuations in $L$ and $\tau$, indicative of 
the tidal influences of the most closely neighboring environs.  
Also in systems with significant substructure, small-scale motions in
non-linear regions and torques caused by nearby structure can effect
large changes in internal angular momentum (BE87).  We see this in a few
of the more massive of the galaxies with noisy $L(t\ga t_{C})$,
where small satellites with eccentric orbits efficiently transfer
angular momentum out of the central system to the surrounding matter.
However, even when the distributions are non-linear at early times, we
still find exceptionally good agreement with the qualitative
predictions of linear-growth theory.

\subsection{Quantitative comparison with linear theory \label{sec5-2-3}}

The predictive capacity of linear theory of tidal torques was
partially examined by BE87, who computed the predicted final value by
measuring the initial $L_{i}$ for each galaxy and evolving it forward
to $a=3$ using the linear time factor such that
\mbox{$L_{pred}=a^{2}\dot{D}L_{i}$}.  They then compared this
semi-empirical prediction with the actual final angular momentum, and
found that the predicted final value tends to be a factor of $\sim 3$
larger than the actual $L_{f}$ with a total scatter about the mean of
roughly the same factor.  
This method failed to fully test the linear theory, since it
improperly assigned the same turnaround epoch to every
galaxy, and relied on the actual initial spin, rather than the
rotation predicted from the external tidal field. 

To properly test the linear-theory predictions, we evaluate the
initial conditions using the initial tidal and inertia tensors, as
prescribed in \mbox{\S \ref{sec4-4}} and 
directly compare the resulting values of \Lone [equation
(\ref{eqn2-5})] to the actual total angular momentum at each galaxy's
individual turnaround and caustic crossing epochs, as well as to the
final value $L_{f}$.  Since the later-time decay of angular momentum
can range from slightly noisy for the SE catalog to strongly varying
in the INT catalog (discussed in the previous subsection), the value
of the angular momentum at the final epoch $L(t_{f})$ may prove
sensitive to which epoch we chose to label as ``final.''  We use two
methods to approximate the final value $L_{f}$ for a smooth decay 
of angular momentum.  We extrapolate a linear least-squares fit for
$L(t>t_{C})$ to find $L_{f}^{\rm lin}$, and also smooth the angular
momentum over time using a 16-point Savitzky-Golay filter 
(Press \etal 1992) to find $L_{f}^{\rm sm}$.   This algorithm does
not require symmetric smoothing, hence its applicability.

\begin{figure} 
 \epsfig{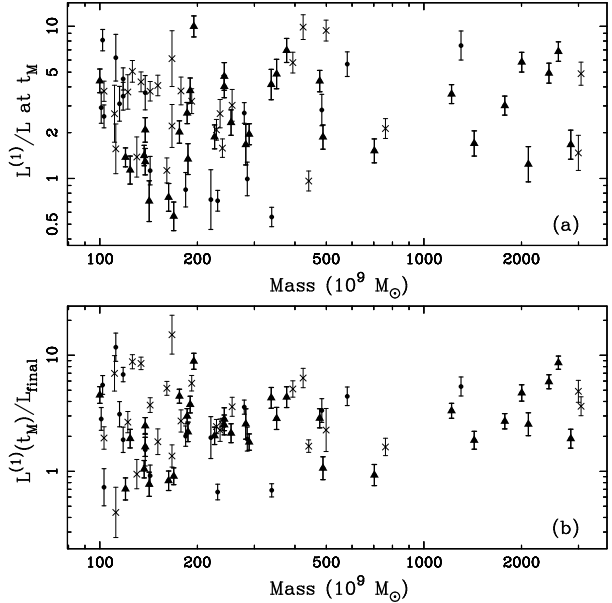}
 \caption{(a) Predicted versus actual total angular momentum at
 turnaround.  (b)  Predicted angular momentum at turnaround versus
 final actual angular momentum.  Both distributions are centred about
 $\sim 3$ with a $\sim 70 $ percent dispersion.  
 Crosses denote the excluded (INT) catalog, triangles indicate members
 of the LG catalog {\em only}, and circles indicate members of the SE
 catalog (LG implicit).}
 \label{lratio}
\end{figure}

\begin{figure} 
 \epsfig{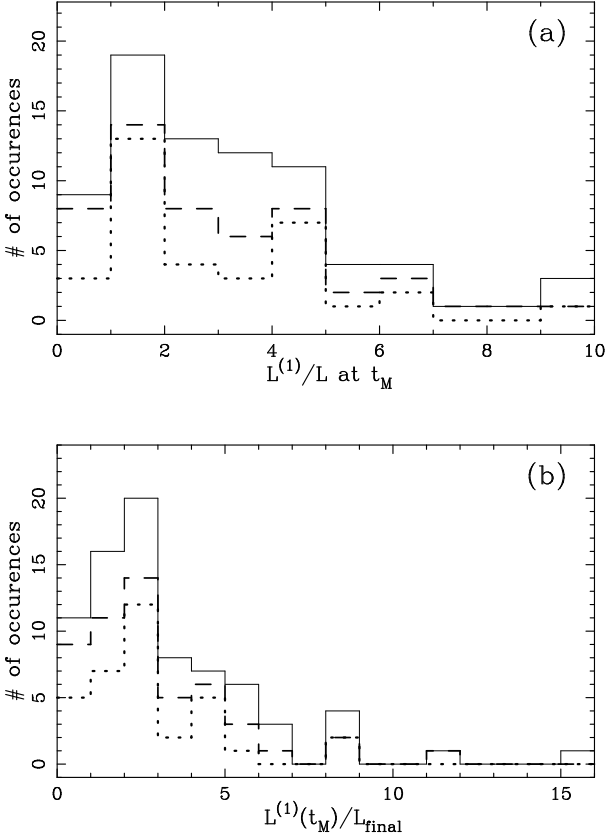}
 \caption{Binned histograms corresponding to ratios plotted in Fig.\
 \ref{lratio}.  The solid curve denotes all galaxies, the dashed curve
 denotes galaxies with $M \ge 200$ and the dotted curve those galaxies
 with $M \ge 1000$ particles. }
\label{lhist}
\end{figure}

In Figure \ref{lratio} we plot galaxy mass versus the ratio of the
predicted to actual angular momentum for the full catalog, using
different symbols for the LG, SE and INT galaxies.  Corresponding
histograms for these ratios are shown in Figure
\ref{lhist}.  We first address the value of predicted to actual
angular momentum at turnaround (panel [a] of both Figures). The
distribution of $L^{(1)}/L$ at $t_{M}$ peaks at a ratio of 
$\sim 1-2$ with a significant tail to larger values.  $L^{(1)}(t_{M})$
generally overestimates the actual value $L(t_{M})$ for a
given galaxy by $\sim 3-4$ with a dispersion of roughly $50-70 $ percent.
BE87 also found the same degree of overestimation with a larger
scatter about the mean.  Since caustic  
crossing, rather than turnaround, is the epoch at which angular
momentum ceases to grow, we also test the correlation of $L^{(1)}/L$
at $t_{C}$.  The distribution appears similar to those in Figs.\
\ref{lratio}, and \ref{lhist} yet with a mean ratio roughly 1 larger
than, and dispersion equal to, that for turnaround. 
We previously noted that the growth of $L(t_{M}<t<t_{C})$ deviates
slightly from linearity.  However, in general $L(t_{C}) \sim
(2-3)L(t_{M})$.  Considering that \Lone grows linearly and also that
$t_{C} \simeq 2t_{M}$, we expect the average value of $L^{(1)}/L$ to be
systematically larger at $t_{C}$ than at $t_{M}$ yet with a roughly
comparable dispersion.

Angular momentum is approximately conserved in the post-turnaround
era, indicating that $L(t_{M})$ should correlate closely 
with $L_{f}$.  If the linear-theory prediction of $L^{(1)}$ at
turnaround does not correlate with the final angular momentum of an
evolved galaxy, then the applicability of the linear theory is
drastically reduced.  We show the ratio of the predicted angular
momentum at turnaround to the final actual angular momentum
$L^{(1)}(t_{M})/L_{f}$ in panel [b] of Figures
\ref{lratio} and \ref{lhist}. The distribution is highly peaked about 
$\sim 2-3$, again with a significant tail to higher values, but with
more power in the lesser values as well.  Given the
typical stochastic decay of $L(t>t_{C})$ seen in simulations, we
expect both the mean (since $L^{(1)}(t_{M})> L(t_{M}) >L_{f}$) and
variance (since the decay rate is different for each galaxy) of this
distribution to be higher than when comparing $L^{(1)}/L$ at $t_{M}$.
In fact, we find that the mean and dispersion for both ratios are
nearly identical.  We find similar results by replacing $L_{f}$ with
$L_{f}^{\rm lin}$ and $L_{f}^{\rm sm}$.  If one approximates the decay
of $L$ as an inherently smooth function of time, then the tidal noise
superposed on that decay is generally only of the order $\la
10 $ percent.  Even for galaxies with $50$ percent variation in
$L(t)$, the linear fit and smoothed value of $L_{f}$ typically differ
from the true final value by $\la 15 $ percent.  Accordingly, the
average ratios for these three methods differ by roughly the same
factor. We conclude that $L_{f}^{\rm lin}$ and $L_{f}^{\rm sm}$ are
consistent methods of approximating the final angular momentum.  There
is little gained in using these two methods to determine $L_{f}$. 
However, the smoothed angular momentum damps out large variations in
tidally disturbed galaxies and thereby allows a more consistent
estimation of the underlying angular momentum evolution at late times,
while the linear fit permits one to extrapolate beyond the limits of
the simulation. 

Provided the measured $L$ did not decay at later times, one would
expect the final 
value $L_{f}$ to be roughly equal to the angular momentum at caustic
crossing, as seen in Galaxy \#22 (Figure \ref{Loft1}[b]).  We
examine the ratio of $L^{(1)}(t_{C})/L_{f}$, and find a peak value of
$\sim (4-5)$, again with a dispersion of $50-70 $ percent.  The mean
is  again larger since $L^{(1)}(t_{C}) > L^{(1)}(t_{M})$.  While
caustic crossing may prove itself more useful than turnaround in
linear theory of collapsing ellipsoidal distributions, it does not
provide any increase in accuracy and cannot be properly implemented
until we efficiently circumvent the numerical decay of angular
momentum.    

\begin{table*}
\caption{Average values for $L^{(1)}/L$}
\label{tbl5-5}
\begin{tabular}{c l l l l l l l l}
   & {$M >$} & {$N$} &
  {$\frac{L^{(1)}(t_{M})}{L(t_{M})}$} &
  {$\frac{L^{(1)}(t_{C})}{L(t_{C})}$} &
  {$\frac{L^{(1)}(t_{M})}{L(t_{f})}$} &
  {$\frac{L^{(1)}(t_{M})}{L^{\rm lin}(t_{f})}$} &
  {$\frac{L^{(1)}(t_{M})}{L^{\rm sm}(t_{f})}$} &
  {$\frac{L^{(1)}(t_{C})}{L(t_{f})}$} 
\\ \hline
Full&     0 & 77 & 2.85 (0.70) & 3.87 (0.74) & 3.10 (0.70)
                 & 2.69 (0.72) & 2.89 (0.75) & 5.73 (0.75) \\
    &   200 & 39 & 3.17 (0.71) & 4.25 (0.61) & 3.11 (0.62)
                 & 2.87 (0.59) & 3.00 (0.64) & 5.73 (0.62) \\
    &  1000 & 11 & 3.79 (0.62) & 4.98 (0.48) & 4.47 (0.54)
                 & 4.29 (0.54) & 4.41 (0.54) & 8.18 (0.45) \\
LG  &     0 & 52 & 2.58 (0.73) & 3.61 (0.73) & 2.72 (0.72)
                 & 2.50 (0.72) & 2.54 (0.73) & 5.08 (0.75) \\
    &   200 & 28 & 2.97 (0.70) & 4.04 (0.66) & 2.90 (0.67) 
                 & 2.72 (0.63) & 2.83 (0.66) & 5.26 (0.67) \\
    &  1000 &  9 & 3.97 (0.62) & 5.04 (0.49) & 4.32 (0.60)
                 & 4.11 (0.59) & 4.34 (0.59) & 8.12 (0.48) \\
SE  &     0 & 34 & 2.61 (0.71) & 3.61 (0.67) & 2.73 (0.69)
                 & 2.51 (0.68) & 2.60 (0.70) & 3.51 (0.70) \\
    &   200 & 20 & 3.31 (0.56) & 4.49 (0.45) & 3.17 (0.60)
                 & 2.86 (0.57) & 3.00 (0.59) & 5.45 (0.52) \\
    &  1000 &  8 & 3.69 (0.63) & 4.17 (0.63) & 4.17 (0.63)
                 & 3.91 (0.60) & 4.06 (0.62) & 6.87 (0.47) \\
INT &     0 & 25 & 3.51 (0.61) & 4.81 (0.76) & 4.05 (0.59)
                 & 3.26 (0.68) & 3.80 (0.72) & 7.87 (0.66) \\
    &   200 & 11 & 3.66 (0.74) & 5.03 (0.43) & 3.60 (0.48)
                 & 3.26 (0.48) & 3.39 (0.62) & 6.95 (0.45) \\
    &  1000 &  2 & 3.20 ($\cdots$) & 4.32 ($\cdots$) & 5.17 ($\cdots$)
                 & 5.22 ($\cdots$) & 4.77 ($\cdots$) & 8.88 ($\cdots$) \\
\hline
\end{tabular}

\medskip
{Averages are listed with the respective fractional
   dispersions in parentheses.} 
\end{table*}

Averages for the six ratios discussed and all four galaxy catalogs
are listed in Table \ref{tbl5-5}.  We further discriminate between
galaxies with masses $M>200$ and $M>1000$ particles for each catalog,
in the event that correlations are mass-dependent.  Each average ratio
is listed with the percent dispersion about the mean. 
The deviation about the mean is higher when including the lowest mass
($M<200$) objects since these have poorer numerical resolution.  
We also find that the average ratios for subsets of galaxies with
$M>1000$ are typically larger than subsets also containing lighter
galaxies ($M>200$ and $M>0$).  As seen in Figure 
\ref{lratio}, less massive objects exhibit stronger scatter 
toward the smaller end of the ratio scale, which lowers the
average and raises the dispersion.  The distributions of these points
have roughly the same peaks and high-end tails for the different
catalogs (Fig.\ \ref{lhist}).  Thus the average and scatter about the
mean are generally comparable 
between catalogs for the same mass scale.  
Except for the mean ratios at caustic crossing, which have
systematically larger values, the mean and dispersion values for
all the tests are highly consistent, indicating that \Lone is well
correlated with both $L(t_{M})$ and $L_{f}$.  

Previous studies (White 1984, BE87) also conclude that linear theory
overestimates $L$ by a factor of $\sim 3$.  We find a scatter about this
mean of roughly $70$ percent for low-mass objects and $50$ percent
as the mass 
scale is increased.  Since our largest objects show stronger
correlations, we expect that improvements in resolution will 
further tighten the correlation between actual and predicted angular
momentum.  

\subsection{Peak heights \label{sec5-3}}

Following the biased-galaxy formation scheme (Kaiser 1984; Politzer \&
Wise 1984; Peacock \& Heavens 1985; Bardeen et al. 1986), we expect
that galaxy formation only occurs for density peaks above a given
threshold of the initial Gaussian density field.  The relative
peak height above the underlying fluctuation of the density field 
is characterized by $\nu = \delta/\sigma$, where $\sigma$ is
determined for the mass scale of the galaxy in question. 
Since evolving peaks are subject to strong tidal fields, not every
high-$\nu$ peak is guaranteed to evolve into a galaxy.  Rather, peaks
can be stretched or sheared such that the highest-$\nu$ regions fail
to undergo infall while lower-$\nu$ peaks collapse into bound
structures (Van de Weygaert \& Babul 1994; EL95).  Rather than assume
that all our galaxies originated from the highest-$\nu$ peaks, we have
the luxury of tracing back the final bound structures to the initial
density field to test the correlation of peak height with collapse.

\begin{figure} 
 \epsfig{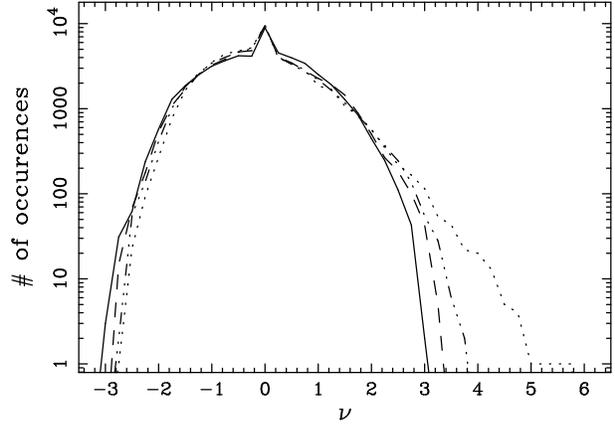}
 \caption{Binned histogram of peak height $\nu$ for $50,\!000$
 randomly placed spheres containing 2500 (solid curve), 1000 (dashed
 curve), 500 (dot-dashed curve) and 100 (dotted curve)
 particles.  }
\label{nuhist}
\end{figure}

For a CDM power spectrum, as the mass scale increases, the
deviation $\sigma$ will decrease since larger volumes more closely
approximate the background density.  Furthermore, low peaks on a
high mass scale become higher-mass peaks when measured on lower
mass scales.  We expect that as the mass scale increases, both the
maximum peak height and the frequency of occurrence of the highest
peaks will decrease.  We tested this by measuring $\nu$ at $5\times
10^{4}$ randomly placed points for mass scales of 2500, 1000, 500 and
100 particles. We found $\sigma = 6.95\times10^{-2},\,
8.68\times10^{-2},\, 0.103,\, 0.145$ respectively, indicating that
$\sigma$ is anticorrelated with the mass scale $M$.  
Histograms for these essays are plotted in Figure \ref{nuhist}.  All
four mass scales have nearly the same distribution for $\nu < 2$.  As
the mass scale decreases, the maximum peak-height increases as well as
the number of high-$\nu$ peaks encountered, as predicted.

From these findings, one might be led to conclude that $\nu$ should
vary inversely with mass scale, such that more low-mass galaxies
should originate from high-$\nu$ peaks.
There are two effects which lead to the opposite
correlation.  First, low-$\nu$ peaks on high mass scales are likely to
contain, and to be identified with, a galaxy arising from a higher-$\nu$
peak on a smaller mass scale.  Second, the high-$\nu$ peaks on small
mass scales evolve quickly and become lower-$\nu$ peaks on higher
mass scales.  Both effects tend toward medium-$\nu$ peaks on medium
mass scales, and a deficit of low-$\nu$, high-mass peaks as well as
high-$\nu$, low-mass peaks.  That implies a correlation of higher
peak height with mass scale.  

\begin{figure}
 \epsfig{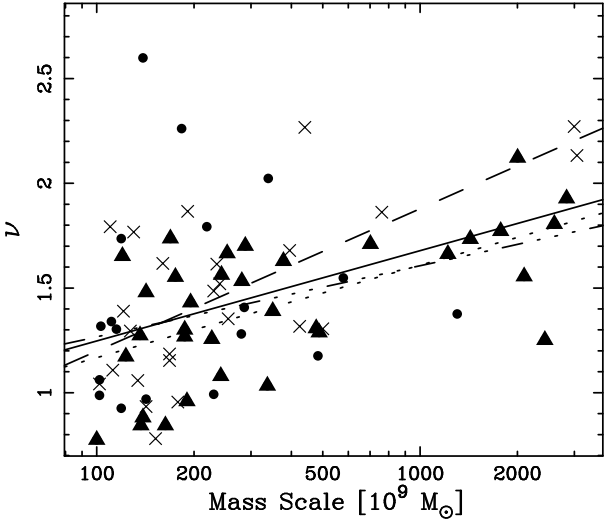}
 \caption{Peak height $\nu$ versus mass scale for the same catalogs as
 Fig.\ \ref{lratio}.  Also plotted are the least-squares linear
 interpolations for the full (solid curve), LG (dot-dot-dot-dashed
 curve), SE (dotted curve) and INT (dashed curve) catalogs.}
 \label{nuvsm}
\end{figure}

\begin{figure} 
 \epsfig{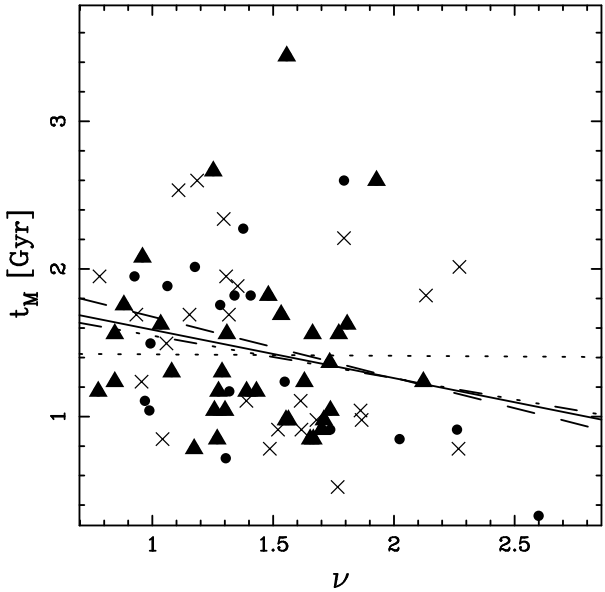}
 \caption{As in Fig.\ \ref{nuvsm} but plotting turnaround time versus
 peak height $\nu$.}
 \label{nuvstm}
\end{figure}

We find values of $\nu$ between 0.7 and 2.6 for the full catalog, with
the lowest and highest peaks being the rarest. In Figure \ref{nuvsm},
we plot the distribution of $\nu$ versus mass scale for the four
catalogs, accompanied by the best-fit lines for each catalog.
On average, $\nu$ is larger for higher-mass
galaxies and there are relatively few high-$\nu$, low-mass 
galaxies, as postulated above.  Furthermore, the distribution of
moderate-mass objects is spread over a wide range of values of $\nu$,
with very few peaks $>2$.  This may also be a consequence of the
``migration'' scenario described above, which makes low-$\nu$,
high-$M$ and high-$\nu$, low-$M$ galaxies comparatively rare. This   
underlines that biasing is a function of scale.   For a high-mass
distribution to evolve into a high-mass galaxy, it appears that 
it must come from a larger-peak density region, otherwise the outer
mass shells do not collapse and the object evolves into a lower-mass
galaxy.  

\begin{figure} 
 \epsfig{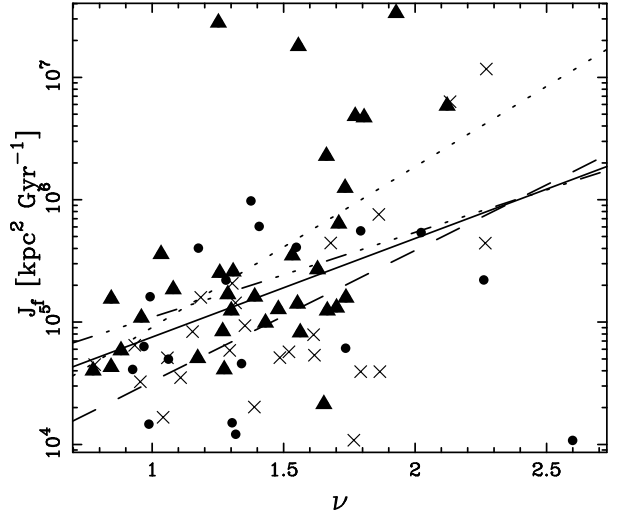}
 \caption{As in Fig.\ \ref{nuvsm} but plotting specific angular
 momentum versus peak height $\nu$. }
\label{jvsnu}
\end{figure}

From equation (\ref{eqn4-5}), we find the turnaround time for a galaxy
scales as $t_{M}\propto \delta_{i}^{-2/3}$.  As $\nu$ increases, the
time a galaxy spends in the linear regime decreases.  We plot $\nu$
versus turnaround time in Figure \ref{nuvstm}, in which we have also
plotted the best-fit lines for all four catalogs.  In all cases, 
$t_{M}$ appears to vary inversely with $\nu$, as expected.  Since a
galaxy acquires most of its angular momentum through tidal
interactions before turnaround, this implies that higher-$\nu$
objects should gain less angular momentum (Hoffman 1986).  We plot
the specific angular momentum $J \equiv L/M$ versus $\nu$ in Figure
\ref{jvsnu}, again complemented with best-fit lines for the four
catalogs.  Instead, we find the moderate trend that angular momentum
increases with 
increasing peak height, most drastically for the SE galaxies.  
If one considers the peaks scenario, higher peaks tend to
be more strongly clustered (Kaiser 1984; Bardeen \etal 1986), implying
that high peaks evolve in environments within stronger local tidal
fields.  Even though these objects spend less time in the linear
regime, the local tidal field must have a more significant effect.
CT96 also find, from angular-momentum probability distribution
functions derived from linear theory, that higher-$\nu$ peaks tend to
having higher $L$. 

\begin{figure}
 \epsfig{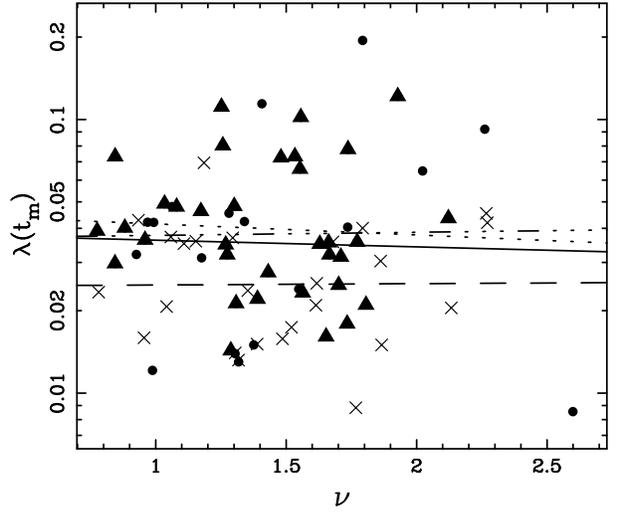}
 \caption{As in Fig.\ \ref{nuvsm} but plotting spin parameter $\lambda$ 
 versus peak height $\nu$. }
\label{lamvsnu}
\end{figure}

A common measurement is the dimensionless spin parameter
$\lambda=LE^{1/2}/GM^{5/2}$.  Due to the problem outlined in
\S\ref{sec5-2-1} in identifying all bound particles within a group,
uncertainty in the particle make-up of a galaxy effects all three
measurables in $\lambda$.  Given this caveat, we examined the spin
parameter measured at turnaround and the final epoch versus peak
height.  It is expected (e.g.\ Hoffman 1988, HP88)
that $\lambda$ and $\nu$ will be anti-correlated 
for galaxies arising from a random density-field. The distributions
are displayed in Figure \ref{lamvsnu}, and 
show no meaningful correlations other than a mean around $\lambda \sim
0.05$ with an appreciable scatter, which is consistent with
previous findings (e.g.,\ BE87, HP88).  Given the
large uncertainties that enter into $\lambda$, this result suffers from
the limited size of our sample.  We will reexamine the
$\lambda-\nu$ anti-correlation with our larger data set in paper II.  

\begin{figure} 
 \epsfig{file=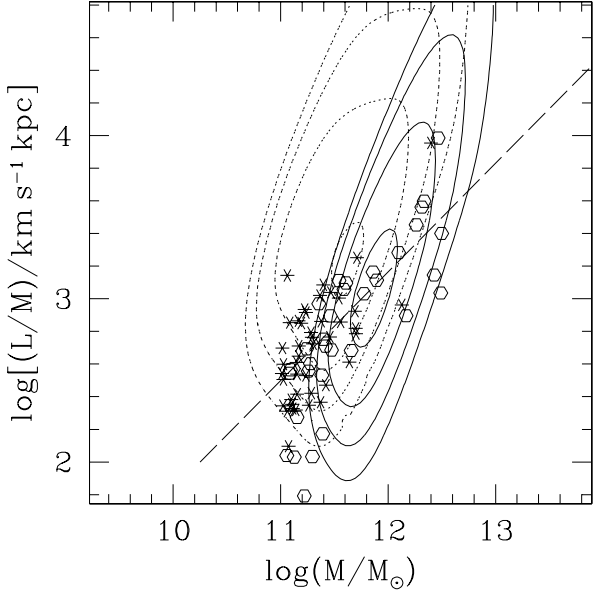,width=80mm}	
 \caption{Mass versus specific angular momentum for the galaxies with
 $\nu < 1.5$ (six-pointed stars) and $\nu \ge 1.5$ (open hexagons),
 superposed on theoretical equiprobability
 contours from Fig.\ 9 of CT96, corresponding to peaks
 with height $1/2 \le \nu \le  3/2$ (dotted contour) and $3/2 \le \nu
 \le 5/2$  (solid contour).  The dashed line indicates the scaling $L/M
 \propto M^{2/3}$.
  }
\label{ctfig9}
\end{figure}

CT96 calculate the equiprobability contours for mass and specific
angular momentum for a CDM power spectrum, onto which they superpose
(their Figure 9) observations by Fall (1983) for the masses and specific
angular momenta of luminous elliptical, Sb and Sc 
galaxies.  Spirals are generally constrained within the
contour limits but do not follow the slope of the probability
distributions.  The elliptical galaxies mostly lie outside the high
probability contours and also exhibit a discrepant slope.  We display
the equiprobability contours for peaks with height $1/2 \le \nu <
3/2$ and $3/2 \le \nu \le 5/2$ from CT96, superposed with data from
our galaxy catalog, separated into galaxies with $\nu < 1.5$
and $\nu \ge 1.5$, in Figure \ref{ctfig9}.
Both sets of galaxies are very well constrained within the
corresponding contours, although the locus of points for $\nu < 1.5$
lies off-centre from the probability maximum.  
Since the probability distribution functions of CT96 have been
derived from equation (\ref{eqn2-5}), this excellent match 
lends validity to the use of ensemble averages derived from linear
theory. 

\subsection{Scalings \label{sec5-4}}

In the previous section, we found that the final value of the
angular momentum $L_{f}$ roughly equals that at turnaround, as
predicted by linear theory.  This value can be estimated from equation 
(\ref{eqn2-5}) using the turnaround condition $\delta=1.07$ [equation
(\ref{eqn2-12})].  This condition can be written equivalently as 
$\delta=-D(t_{M})\nabla^{2}\psi$, from which the previous  
relation yields the equivalent condition $D_{M} \simeq 1/\nabla^{2}\psi$
where $D_{M}=D(t_{M})$.  
Denoting the initial mass and radius of the protogalaxy by $M$ and
$R_{0}$ respectively, we insert this into equation (\ref{eqn2-5}) (White
1994; CT96), 
\begin{eqnarray} L_{f} & \sim & a^{2}_{M} \dot{D_{M}} \nabla^{2} \psi
  MR^{2} = a^{2}_{M} \frac{\dot{D_{M}}}{D_{M}} MR^{2}_{M} \nonumber \\
  & \propto & 
  \frac{\dot{D_{M}}}{D_{M}} \rho^{-2/3}_{b,M} M^{5/3} \nonumber \\
  & \propto & \left( \frac{a_{M}\dot{D_{M}}}{\dot{a_{M}}D_{M}} \right)
    \frac{\dot{a_{M}}}{a_{M}} \rho^{-2/3}_{b,M} M^{5/3} \nonumber \\
  & \propto & \Omega^{0.6} H(\Omega H^{2})^{-2/3} M^{5/3} \nonumber \\ 
  & \propto & \Omega^{-0.07} \left( \frac{\dot{a_M}}{a_{M}}
  \right)^{-1/3} M^{5/3}, \label{eqn5-2}
\end{eqnarray}
where in the last step we have used the approximation given by Peebles
[1993, eq.\ (5.120)].  For an Einstein-de Sitter universe, $a \propto
t^{2/3}$ and $\Omega=1$ and the previous expression simplifies to 
\begin{equation} 
   L_{f} \cong L(t_{M}) \propto t_{M}^{1/3} M^{5/3}. \label{eqn5-3}
\end{equation}
From a physical standpoint, we expect objects with larger masses to
gain a greater final angular momentum, not only due to the linear mass
term in the basic expression for $L(t)$, but also since larger masses
tend to fill larger volumes, whereby more massive galaxies will have
larger moment arms on which the tidal field will act.  
Additionally, the longer the tidal field acts upon an object, the
greater the acquired angular momentum, hence the dependence on 
the time at which angular momentum stops growing.  However, since the
turnaround time is a function of overdensity (hence mass), the time
dependence in equation \ref{eqn5-3} is not fully given by $t_{M}^{1/3}$. 

\begin{figure*} 
 \epsfig{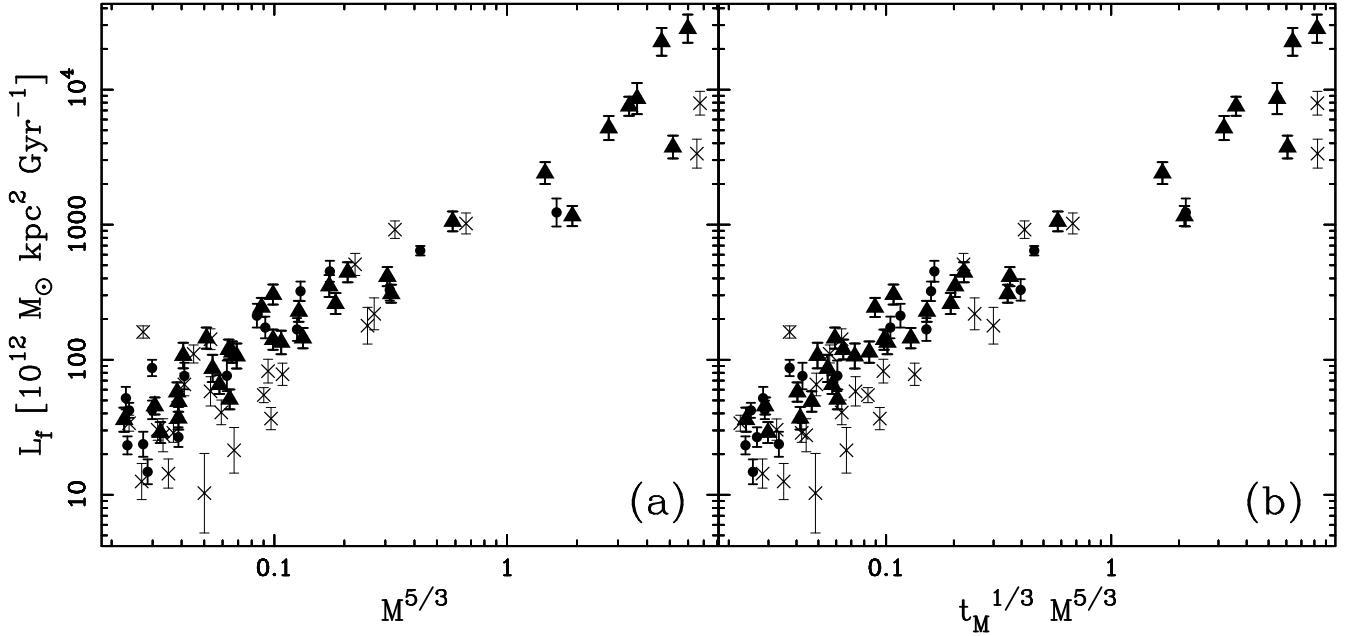}
 \caption{Final angular momentum for the same catalogs as Fig.\
 \ref{lratio} versus scalings determined from eq.\ (\ref{eqn2-5}).  (a)
 $L_{f} \propto M^{5/3}$.  (b)  $L_{f} \propto t_{M}^{1/3}M^{5/3}$
 (the best $\chi^{2}$ residual for all catalogs).
 }
\label{lmtnu}
\end{figure*}

We test equation (\ref{eqn5-3}) directly in Figure \ref{lmtnu}, in
which we have 
plotted $L_{f}$ versus $M^{5/3}$ and $t_{M}^{1/3} M^{5/3}$ 
for all four galaxy catalogs.  With a 
correct scaling, the distribution of points will be linear.  We find
that the scaling $L_{f} \propto t_{M}^{1/3} M^{5/3}$ yields the
smallest $\chi^{2}$ residual and also has the most linear distribution
in all four catalogs. 

\begin{table*}
\begin{minipage}{110mm}
\caption{Scaling parameters for SVD fitting}
\label{tbl5-6}
\begin{tabular}{c l l c c c c c c c c c c}  
  &  &  &
 \multicolumn{2}{c}{$M^{\alpha}$} &  &
 \multicolumn{3}{c}{$M^{\alpha} t_{M}^{\beta}$} &  &
 \multicolumn{3}{c}{$M^{\alpha} t_{C}^{\beta}$} 
 \\
  & {$M>$} & {$N$} &
 {$\alpha$} & {$\chi^{2}$} &  &
 {$\alpha$} & {$\beta$} & {$\chi^{2}$} &  &
 {$\alpha$} & {$\beta$} & {$\chi^{2}$}  
\\ \hline
Full&   0 &77&1.76 & 2.63 && 1.67 & 0.645 & 2.21 && 1.71 & 0.595 & 2.36 \\
    & 200 &39&1.72 & 2.26 && 1.63 & 0.481 & 2.10 && 1.64 & 0.394 & 2.18 \\
LG  &   0 &52&1.76 & 1.55 && 1.66 & 0.595 & 1.22 && 1.70 & 0.498 & 1.38 \\
    & 200 &28&1.67 & 1.57 && 1.59 & 0.446 & 1.43 && 1.61 & 0.295 & 1.53 \\
SE  &   0 &34&1.76 & 1.43 && 1.64 & 0.772 & 1.09 && 1.70 & 0.523 & 1.28 \\
    & 200 &20&1.75 & 1.72 && 1.47 & 1.09  & 1.22 && 1.55 & 0.821 & 1.49 \\
\hline
\end{tabular}
\end{minipage}
\end{table*}

To more properly test the validity of these scalings, we must see if
they naturally occur in the data.  We use the singular value
decomposition (SVD) method of generalized least-squares fitting (Press \etal
1992) to solve for the power-law exponents by rewriting equation
(\ref{eqn5-3}) as a linear combination of logarithms
\begin{equation}
 \log{L_{f}}=\log({\rm const}) + \alpha\log{M} + \beta\log{t_{M}} 
 \label{eqn5-4}  
\end{equation}
Since the growth of angular
momentum continues past turnaround and up until crossing time
$t_{C}$, we fit equation (\ref{eqn5-4}) for $t_{C}$ as well as $t_{M}$.
Selected results are listed in Table \ref{tbl5-6}.  We find that
$L_{f}$ scales with mass $M^{1.5-1.75}$, which is highly consistent
with an $M^{5/3}$ dependence.  The scalings in turnaround and caustic
crossing time both vary inconsistently, and differ from the $t^{1/3}$
dependence, as expected from our discussion above.  At present, 
our findings remain inconclusive regarding the
temporal factor in equation (\ref{eqn5-3}).  Considering that our
galaxies cover more than a decade in mass, whereas the turnaround
times typically differ by less than 40 percent, it is not surprising
that $L_{f}$ scales predominantly on the mass.  It is perhaps more
surprising that the empirically determined mass scaling so closely
matches linear theory.   

\subsection{Correlations \label{sec5-5}}

\begin{figure} 
 \epsfig{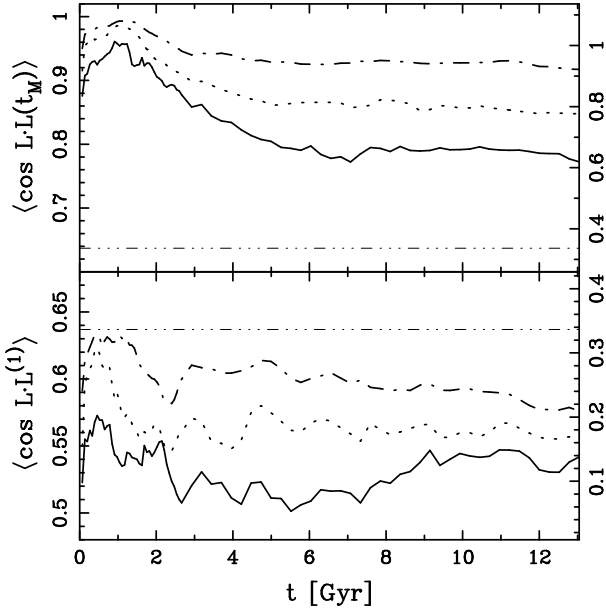}
 \caption{Average inner products of spin-vectors in time. An
 uncorrelated  distribution will  
 remain at $2/\pi = 0.64$, marked by the dot-dot-dot-dashed line.  
 The right-hand abscissa indicates the fraction of randomly oriented pairs
 of vectors with separation angle less than $\theta=30^{\circ}$ which
 are required to yield the corresponding average inner product on the
 left-hand abscissa.
 Top panel: the mean value of $\cos{|{\bmath L} \cdot {\bmath
 L}(t_{M})|}$ for the full catalog (solid line), LG catalog (dotted
 line) and LG catalog (dot-dashed line). Bottom panel: 
 the mean value of $\cos{|{\bmath L} \cdot {\bmath L}^{(1)}|}$ for the
 full, SE and LG galaxy catalogs versus time.  
  }
 \label{ll1corr}
\end{figure}

Our final examination of linear theory is to check whether equation
(\ref{eqn2-5}) 
predicts the correct spin axis for a galaxy.  Given the effects of
the local tidal field examined in \S \ref{sec5-2-2}, it is possible
that the direction of ${\bmath L}$ will vary over time.  In this case,
since the direction of the predicted ${\bmath L}^{(1)}$ remains fixed at
its initial value, we find little reason to expect any long-term
correlation between the two vectors.  To test the stability of
${\bmath L}(t)$, we plot the average value of the inner product of the
angular momentum at turnaround ${\bmath L}_{M}$ with ${\bmath L}(t)$
for the full, LG and SE catalogs in Figure \ref{ll1corr} (top panel).
Following turnaround, which occurs at roughly 2 Gyr, angular momentum
is relatively stable in time, and hence the local tidal field
predominantly affects only the magnitude rather than the direction.

To quantify the allignment of two vectors, we consider them weakly
parallel if the angle between them is less than $30^{\circ}$.
Table \ref{tbl5-7} lists the number of galaxies in each catalog
and mass scale for which the two vectors ${\bmath L}$ and ${\bmath L}^{(1)}$
are weakly parallel or anti-parallel.  We consider the vectors
correlated if they satisfy the weakly parallel condition for at least
90 percent of the epochs in the simulation. The probability
of two randomly oriented vectors are separated by $\theta$ is given
by the ratio of the arc subtended by $\theta$ to that of the unit 
circle $p=\theta/\pi$.  For $\theta=30^{\circ}$, we expect $1/6$ of
randomly placed pairs of vectors to be aligned. The
last column of Table \ref{tbl5-7} 
lists the number of correlated galaxies from each subset we expect if
the orientation of ${\bmath L}$ and ${\bmath L}^{(1)}$ is completely random.
In no case is the correlation more probable than random. 
To confirm this finding, we plot in Figure \ref{ll1corr} (bottom
panel) the average absolute-value
inner product of the two vectors for the full, LG and SE catalogs in
time.  For a randomly orientated distribution, 
$\langle {\bmath L} \cdot {\bmath L}^{(1)}\rangle = 2/\pi \simeq 0.64
$, which we have marked in both panels.  In the right-hand abscissa of
each panel, we have indicated the fraction of randomly oriented pairs
of vectors with separation angle less than $\theta=30^{\circ}$ which
are required to yield the corresponding average inner product on the
left-hand abscissa.  We see that the alignment between the actual and
predicted angular momentum does not occur more often than random from
the above probability argument, and only occurs for roughly 10\% of
the galaxies in the full catalog, 18\% of the LG, and $\sim$24\% of
the SE catalog. It thus appears that linear theory predicts the
correct spin axis with, at most, only marginal statistical
significance.   

\begin{table}
\caption{Correlation statistics for ${\bmath L} \| {\bmath L}^{(1)}$}
\label{tbl5-7}
\begin{tabular}{c l l c c l}
   & {$M>$} & {$N$} &
  {$N_{\rm correlated}$} & 
  {$N_{\rm anti-corr.}$} &
  {$N_{\rm random}$}
\\ \hline
Full&    0 & 77 & 5 & 4 & 12.8  \\
    &  200 & 39 & 3 & 3 & 6.5  \\
    & 1000 & 11 & 1 & 2 & 1.8 \\
LG  &    0 & 52 & 4 & 4 & 8.7  \\
    &  200 & 28 & 2 & 3 & 4.7  \\
    & 1000 &  9 & 1 & 2 & 1.5 \\
SE  &    0 & 34 & 4 & 4 & 5.7  \\
    &  200 & 20 & 2 & 3 & 3.3  \\
    & 1000 &  8 & 1 & 2 & 1.3 \\
\hline
\end{tabular}
\end{table}

\section{Discussion and conclusions \label{sec6}}

In this paper, we have examined the predictions of linear collapse and
linear tidal-torque theory within the framework of a high-resolution
$N$-body and hydrodynamical
simulation.  This permits the study of evolution well into the
non-linear regime, thereby placing important constraints on the
validity and applicability of linear theory to the full range of
evolutionary periods.  In this paper, we have examined the
evolution of dark matter, typical of galactic dark halos.  

The simplest treatment of galaxy evolution at early times follows the
spherical-collapse model, in which we treat the proto-galaxy as a
spherically isotropic closed (\ie overdense) distribution embedded
within an Einstein-de Sitter universe.  The observed evolution is
qualitatively described by the ellipsoidal-collapse model (Peebles
1980; EL95) in which the short axis of the rotating ellipsoidal 
proto-galaxy first collapses into a pancake, followed by collapse of
the secondary and tertiary axes which form caustics that relax into
virialized triaxial systems.  
While the spherical-collapse model greatly oversimplifies the 
observed collapse of galaxies (namely, it is a hierarchical and
ellipsoidal process), analytic predictions correlate strongly with
empirical measurements.  The turnaround time $t_{M}$ corresponds  
closely with the initial collapse of the short ellipsoidal axis.
Furthermore, the epoch $t_{C}$ at which this spherical model predicts
shell-crossings correlates closely with the collapse of the long axes
of the evolving proto-galaxy.  

The linear-theory tidal-torque formalism (\S2) predicts that
the angular momentum $L$ of a galaxy will grow linearly [equation
(\ref{eqn2-5})] 
until turnaround, at which point self-gravitation dominates
large-scale tidal torque.  Since the quadrupole moment (on which the
tidal field acts) of an elliptical ensemble remains significant until
the long axis collapses, we  synthesize a ``standard evolution'' in
which $L(t)$ grows linearly beyond turnaround, turns over during shell
crossing, and remains constant as the galaxy relaxes and
virializes.  As {\em local} tidal-interactions increase in strength and
frequency, the evolution of $L$ becomes increasingly noisy, however
all our galaxies follow an underlying linear growth of $L$ at early
times.  This agrees with analytic non-linear perturbative
solutions of the Lagrangian fluid equations (Catelan \& Theuns, 1996b)
which indicate that the initial 
torque on a galaxy is a good estimate of the torque during the entire
period of angular-momentum acquisition, i.e.\ angular momentum 
will grow linearly in time since tidal torque is approximately
constant.  Post shell-crossing, the measured $L$ appears to
decrease, as a numerical result of the redistribution of angular
momentum to bound particles outside the overdensity cutoff generated
during galaxy-identification.  Unless all bound particles are
identified in a galaxy (an $N^{2}$ calculation), the apparent decay of
$L$ at late times is a generic numerical trend of simulations.      

We test the predicted total angular momentum, {$L^{(1)}$, by
comparing the value calculated from equation (\ref{eqn2-5}) to the true
value of $L$ at the turnaround, caustic crossing, and final epochs.
Equation (\ref{eqn2-5}) at turnaround systematically
overestimates the true turnaround $L(t_{M})$ 
and the final $L_{f}$ by a factor of $\sim 3$ with a
dispersion of roughly $70 $ percent for lower-mass, and 50 percent for
higher-mass galaxies.  
This broad distribution of predicted versus actual angular momentum is
expected from statistical analysis (HP88), and can
be easily understood through, e.g.\ 
time-varying tidal effects, ellipsoidal collapse, and uncertainty in
the actual epoch of decoupling of the galaxy from the large-scale
tidal field.  The nearly ubiquitous overestimation of $L$ by linear
theory observed in this and previous work is less easily explained.
In the linear theory formalism, the angular momentum imparted to a
galaxy equals the torque of a large-scale shear field on the moment
arm of the initial particle distribution, which scales as the square
of the distance.  Furthermore,
equation \ref{eqn2-5} explicitly assumes that the Taylor expansion of
the potential $\psi$ is constant both in time and throughout the
collapsing volume at early times.  If this approximation breaks down
for outlying particles in the initial distribution (see Figs.\
\ref{gal12}, \ref{gal22}) the quadratic dependence of angular momentum
with distance will inflate the predicted value {$L^{(1)}$
(A. Heavens \& R. Jimenez 1999, private communication).  
This motivates a study of the inner mass-shells of collapsing galaxies, 
for which we expect linear theory will more correctly model actual
evolution. 

The agreement of turnaround and final angular momenta implies that
$L(t_{M})$ is a robust estimator of $L_{f}$.  However, given the result
that angular momentum grows beyond turnaround due to ellipsoidal
collapse, we suspect that if the decay of $L$ post-shell-crossing can
be eliminated then $L(t_{C})$ at shell-crossing will prove the more
useful quantity.  We find that the predicted direction of the angular
momentum is only nominally more correlated with the true spin axis
than we would expect from purely random alignment.  
Considering the rapid variation of local tidal fields due to neighbors
and the non-linear redistribution of the inertial properties of an
evolving galaxy, it is of little surprise that equation (\ref{eqn2-5})
only weakly predicts the direction of the angular momentum, and of
even greater surprise that it does so closely predicts the magnitude. 
Nonetheless, this underlines the limitations of applying linear theory
to a highly non-linear problem.

Linear theory predicts that the final angular momentum varies
proportionally to $M^{5/3}$ [equation (\ref{eqn5-3})], which we find
to be robust from 
generalized least-squares fitting.
Equations (\ref{eqn2-5}) and (\ref{eqn5-3}) further permit the
calculation of ensemble averages and probability distributions without
the need for lengthy numerical simulations.  The predicted probability
distribution of specific angular momentum versus mass (CT96), for
example, matches closely with our data (\S\ref{sec5-3}).  This tool
may permits the numerical calculation of the angular-momentum function 
using the Press-Schecter formalism (Press \& Schecter 1974), which
estimates the mass function of collapsed objects in the universe.  
Theoretical correlations of shape, morphology, spin, and density
will constrain models of galaxy formation 
once observations of dark-halo properties become available.
It appears, however, that correlations involving the direction of the
angular-momentum vector (such as the misalignment of ${\bmath L}$ and the
minor body axis) must be relegated to numerical simulations, since
linear-theory predictions of the spin axes are statistically
unreliable.   

We examine trends in the initial density peaks of our galaxies, and
compare them to predictions from the Gaussian-peaks formalism.
Higher-mass objects tend to originate from higher initial density
peaks, with relatively few high-mass low-peak or low-mass high-peak
objects.  
We question whether tidal torques not only spin up proto-galaxies
before turnaround, but also influence the volume of mass from which
the galaxy evolves.  Van de Weygaert \& Babul (1994) find that the
evolution of density peaks is highly sensitive to external shear,
which can disrupt one peak into many halos or form halos from
low-$\nu$ environments.  EL95 determine from their
study of elliptical collapse that the geometry of the collapsing
region is determined largely by the external shear, and not uniquely
by the initial high-density peak.  Were there no significant tidal
shear, anisotropies in the primordial medium would evolve roughly
spheroidally, such that the final object would map surjectively to the
initial density peak.  However, local and large-scale tidal shear
distorts the minor anisotropies in the high-density peak, resulting
in a triaxial system which accretes mass from the surrounding
lower-density shells.  The final bound object rarely resembles the
initial region from which it evolved, and cannot be directly mapped
back to the initial density peak.  The spherical-collapse model is
inadequate in this respect, since it cannot quantify the strong role
of tidal shear on the collapse history.  

From the spherical-collapse model, one expects that 
objects evolving from higher peaks collapse more quickly and spend
less time in the linear regime. However, contrary to the corollary of
this trend, galaxies from higher peaks also have higher final angular
momenta.  HP88 found this result for power
spectra with fluctuations larger than the small-scale smoothing
length.  Since high peaks are more statistically correlated, they
experience stronger tidal forces, and hence
galaxies evolving from higher peaks are potentially subject to
stronger local torque.  These galaxies have shorter collapse times
since external tidal shearing tends to accelerate collapse and act as
a source of internal gravity  (Hoffman 1986; Zaroubi \& Hoffman 1993;
Bertschinger \& Jain 1994; EL95). Perhaps more
straightforward is the fact that, from the scaling correlation in
equation (\ref{eqn5-4}), angular momentum scales significantly more
strongly with 
mass than turnaround time.  Although peak height and collapse time
are inversely proportional, the direct variation of mass and
peak height dominates, from which we expect that peak
height and angular momentum will vary proportionally as well.  

Our simulation evolved particles within only
one-eighth of the comoving box.  This significantly decreased the
necessary CPU time while providing a realistic large-scale tidal
field.  However, galaxies evolving at the vacuum boundary of our
subsampled region as well as those containing particles from the third,
supermassive dark-matter species had to be disregarded. 
Paper II will discuss this analysis within the framework of a full-scale,
single-component simulation with $128^{3}$ particles and identical
initial conditions.  This will provide at
least eight times the number of bound objects, span a higher mass
range and will not suffer from conditions which force us to eliminate
objects from our catalog.  Higher mass scales will permit a study of
internal mass shells, for which we 
expect that the consistent overprediction by linear theory of actual
angular momentum will be resolved.  It is uncertain how significantly
the gas component effects the dynamics of dark-halo evolution; 
by comparing identical halos in both simulations, we will also examine
this question. 

Numerous other statistical tests could be performed on our data set.  
Correlation functions of the principal axes and actual angular
momentum of nearby neighbors (Splinter \etal 1997) can verify the
underlying physics of the tidal-torque scenario, since the large-scale
tidal field should cause nearest neighbors to spin about roughly the
same axis, while the most distant neighbors should counter-rotate.  
BE87 test alignment statistics of axis vectors, angular-momentum and
separation vectors between nearest neighbors to also look for
coherence from tidal fields.   The misalignment vector between the
apparent spin axis and the projected short elliptical 
axis may indicate relationships between triaxial systems and
ellipticities, and has been studied by, \eg BE87; Frenk \etal 1988;
Quinn \& Zurek 1988; Warren \etal 1991; Franx, Illingworth \& de Zeeuw
1991; and Dubinski 1992.  
These tests rely on a larger statistical sample with larger dynamical
resolution than we have available, and will thus be addressed in paper
II.  We anticipate that the substantially larger data-set 
will also reduce the uncertainty in the linear-theory
prediction of angular momentum and permit more quantitative 
conclusions regarding peak-height correlations and scalings.

\section*{Acknowledgements}
B.S. wishes to thank Jim Applegate for numerous enlightening
discussions on celestial dynamics and mechanics, and Gino Thomas for
his Socratic cynicism.  Additional thanks to Raul Jimenez, Alan
Heavens, Paolo Catelan and Tom Theuns for their interest, discussions
and ideas, and to our referee, Joshua Barnes.  
This work was supported by the U.S. Department of Energy Outstanding
Junior Investigator Award under contract DE-FG02-92ER40699, NASA grant
NAG5-3091, the Alfred 
P. Sloan Foundation, Columbia Astrophysics Lab and the Columbia
Department of Astronomy.  This is contribution No.\ 676 from the
Columbia Astrophysics Laboratory, and CU-TP-927 from the Columbia
Physics Department.


\begin{thebibliography}{}

\bibitem{} Bardeen J.M., Bond J.R., Kaiser N., Szalay A.S.,
   1986, Ap, 304,15
\bibitem{} Barnes J., Efstathiou G., 1987,  ApJ, 319, 575 (BE87)
\bibitem{} Bertschinger E., Jain B., 1994, ApJ, 431, 486
\bibitem{} Binney J., Silk, J., 1979, MNRAS, 188, 273
\bibitem{} Binney J., de Vaucouleurs, G., 1981, MNRAS, 194, 679
\bibitem{} Bond J.R.,  Efstathiou, G., 1984, ApJ, 285, L45
\bibitem{} Bond J.R.,  Meyers, S.T., 1993, CITA, preprint 93/27
\bibitem{} Bond J.R.,  Szalay, A.S., 1983, ApJ, 274, 443
\bibitem{} Catelan P.,  Theuns T., 1996a,  MNRAS, 282, 436
   (CT96) 
\bibitem{} Catelan P., Theuns T., 1996b, MNRAS, 282, 455
\bibitem{} de Theije P.A.M., Katgert P., van Kampen E., 
   1995,  MNRAS, 273, 30
\bibitem{} Doroshkevich, A.G., 1970,  Afz, 6, 581
\bibitem{} Dubinski J., 1992, ApJ, 401, 441
\bibitem{} Dubinsky J., 1994, ApJ, 431, 617
\bibitem{} Dubinski J.,  Carlberg R.G., 1991,  ApJ, 378, 496
\bibitem{} Efstathiou G., Bond J.R., White S.D.M., 1992,
   MNRAS, 258, 1
\bibitem{} Efstathiou G., Eastwood J.W., 1981,  MNRAS,
  194, 503
\bibitem{} Efstathiou G., Jones B.J.T., 1979,  MNRAS,
   186, 133
\bibitem{} Efstathiou G., Silk J.I., 1983, Fund. Cosm. Phys.,
   9, 1
\bibitem{} Eisenstein D.J., Hut P., 1998,  ApJ, 498, 137
\bibitem{} Eisenstein D.J., Loeb A., 1995,  ApJ, 439, 520 (EL95)
\bibitem{} Evrard A.E., 1988, MNRAS, 235, 911
\bibitem{} Fasano G., Vio R., 1991, MNRAS, 249, 629
\bibitem{} Franx M., Illingworth G., de Zeeuw T., 1991, ApJ,
   383, 112
\bibitem{} Goldstein H., 1980, Classical Mechanics.
   Addison-Wesley Publishing Co., p. 147
\bibitem{} Gunn J.E.,  Gott J.R., 1972,  ApJ, 176, 1
\bibitem{} Heavens A.F., Peacock J.A., 1988,  MNRAS, 232, 339 (HP88)
\bibitem{} Hockney R.W.,  Eastwood J.W., 1981,  Computer
  Simulation Using Particles. McGraw-Hill, New York
\bibitem{} Hoffman Y., 1986, ApJ, 301, 65
\bibitem{} Hoffman Y., 1988, ApJ, 329, 8
\bibitem{} Hoyle F., 1949, in Burgers, J. M., van de Hulst, H. C.,
  eds., in  Problems of Cosmical Aerodynamics, Central Air Documents,
   Dayton, Ohio, p. 195
\bibitem{} Kaiser N., 1984,  ApJ, 284, L9
\bibitem{} Katz N., 1991,  ApJ, 368, 325
\bibitem{} Katz N., White S.D.M., 1993, ApJ, 412, 455
\bibitem{} Kuhlman B., Melott A.L., Shandarin S.F., 1996, ApJL, 470, L41
\bibitem{} Levison H.F.,  Richstone D.O., 1987, ApJ, 314, 476
\bibitem{} Monaghan J.J., 1992  ARAA, 30, 543
\bibitem{} Ostriker J., 1993, ARAA, 31, 689.
\bibitem{} Partridge R.B., Peebles P.J.E., 1967, ApJ,
   147, 868	
\bibitem{} Peacock J.A., Heavens A.F., 1985,  MNRAS, 217, 805
\bibitem{} Peebles P.J.E., 1969,  ApJ,155, 393
\bibitem{} Peebles P.J.E., 1971, Astr. Ap., 11 377
\bibitem{} Peebles P.J.E., 1980,  The Large Scale Structure
  of the Universe. Princeton University Press, Princeton, NJ
\bibitem{} Peebles P.J.E., 1982, ApJ, 263, L1
\bibitem{} Peebles P.J.E., 1993,  Principles of Physical
  Cosmology. Princeton University Press, Princeton, NJ
\bibitem{} Plionis M., Barrow J.D., Frenk C.S., 1991, 
    MNRAS, 249, 662
\bibitem{} Politzer H.D.,  Wise M.B., 1984, ApJ, 285, L1
\bibitem{} Press, W.H., et al.\ 1992, Numerical Recipes in Fortran
   2nd ed. University Press, Cambridge
\bibitem{} Quinn T., Binney J., 1992,  MNRAS, 255, 729
\bibitem{} Ryden B.S. 1988,  ApJ, 329, 589
\bibitem{} Ryden B.S. 1992, ApJ, 393, 445
\bibitem{} Schwarzchild M., 1979, ApJ, 232, 236
\bibitem{} Sciama D.W., 1955,  MNRAS, 115, 3
\bibitem{} Shandarin S.F., Zel'dovich Ya.B., 1989,  Rev Mod
  Phys, 61, 185
\bibitem{} Smoot G.F. et al., 1992, ApJ, 396, L1
\bibitem{} Splinter R.J., Melott A.L., Linn A.M., Buck C.,
  Tinker J., 1997, ApJ, 479, 632
\bibitem{} Statler T.S., 1987, ApJ, 321, 113
\bibitem{} Summers F.J., 1993, PhD Thesis, Univ. of California, Berkeley
\bibitem{} Summers F.J., Davis M.,  Evrard A.E., 1995,  ApJ,
  454, 1
\bibitem{} Thuan T.X.,  Gott J.R., 1977  ApJ, 216, 194
\bibitem{} van de Weygaert R.,  Babul A., 1994, ApJ, 425, L59
\bibitem{} Warren M.S., Quinn P.J., Salmon J.K.,  Zurek W.H., 
  1992, ApJ, 399, 405
\bibitem{} White S.D.M., 1984,  ApJ, 286, 38
\bibitem{} White S.D.M., 1994, astro-ph/9410043
\bibitem{} Zaroubi S., Hoffman Y., 1993, ApJ, 414, 20
\bibitem{} Zaroubi S., Naim A., Hoffman Y., 1996, ApJ, 457, 50
\bibitem{} Zel'dovich Ya.B., 1970,  AA, 5, 84.
\end{thebibliography}
\end{document}